\def\year{2021}\relax
\documentclass[letterpaper]{article} 
\usepackage{aaai21}  
\usepackage{times}  
\usepackage{helvet} 
\usepackage{courier}  
\usepackage[hyphens]{url}  
\usepackage{graphicx} 
\usepackage{natbib}  
\usepackage{caption} 
\usepackage[utf8]{inputenc}
\usepackage[english]{babel}
\usepackage{booktabs}
\usepackage{array}  
\usepackage[skip=0.333\baselineskip]{caption} %
\usepackage{titlesec}
\usepackage[monochrome]{xcolor}
\usepackage{parskip}
\usepackage{centernot}
\usepackage{xcolor}
\usepackage{here}
\usepackage{flushend}
\usepackage{float}
\definecolor{teal}{HTML}{008080}
\usepackage{hyperref}
\usepackage{amssymb}
\hypersetup{
  colorlinks=true,
  citecolor=teal
}
\usepackage{booktabs, tabularx}
\newcolumntype{C}{>{\centering\arraybackslash}X}
\usepackage{graphicx}
\usepackage{multirow}
\usepackage{enumitem}
\usepackage{xspace}
\usepackage{blindtext}
\usepackage[font=bf]{caption}
\usepackage{subcaption}
\usepackage{amsmath,array,graphicx}
\usepackage{multicol}

\usepackage{xcolor,colortbl}
\newcommand{\mc}[2]{\multicolumn{#1}{c}{#2}}
\definecolor{Gray}{gray}{0.85}
\definecolor{LightCyan}{rgb}{0.88,1,1}
\newcommand{\squishlist}{
 \begin{list}{$\bullet$}
  { \setlength{\itemsep}{1pt}
     \setlength{\parsep}{0pt}
     \setlength{\topsep}{0pt}
     \setlength{\partopsep}{20pt}
     \setlength{\leftmargin}{1.5em}
     \setlength{\labelwidth}{1em}
     \setlength{\labelsep}{0.5em}}}

\newcommand{\squishend}{
  \end{list}  }
\AtBeginDocument{%
  \providecommand\BibTeX{{%
    \normalfont B\kern-0.5em{\scshape i\kern-0.25em b}\kern-0.8em\TeX}}}
\begin{document}
\title{\huge A deep dive into the consistently toxic 1\% of Twitter}
\author{\large Hina Qayyum$^\dagger$, Benjamin Zi Hao Zhao$^\dagger$, Ian D. Wood$^\dagger$\\Muhammad Ikram$^\dagger$, Mohamed Ali Kaafar$^\dagger$, Nicolas Kourtellis$^*$\\
$^\dagger$\{hina.qayyum, ben\_zi.zhao, ian.wood, muhammad.ikram, dali.kaafar\}@mq.edu.au, $^*$nicolas.kourtellis@telefonica.com\\
$^\dagger$ Macquarie University, $^*$ Telefonica Research\\
}
\maketitle
\begin{abstract}

Misbehavior in online social networks (OSN) is an ever-growing phenomenon. The research to date tends to focus on the deployment of machine learning to identify and classify types of misbehavior such as bullying, aggression, and racism to name a few. The main goal of identification is to curb natural and mechanical misconduct and make OSNs a safer place for social discourse.
Going beyond past works, we perform a longitudinal study of a large selection of Twitter profiles, which enables us to characterize profiles in terms of how consistently they post highly toxic content. 
Our data spans 14 years of tweets from 122K Twitter profiles and more than 293M tweets. 
From this data, we selected the most extreme profiles in terms of consistency of toxic content and examined their 
tweet texts, and the domains, hashtags, and URLs they shared. 
We found that these selected profiles keep to a narrow theme with lower diversity in hashtags, URLs, and domains,
they are thematically similar to each other (in a coordinated manner, if not through intent), 
and have a high likelihood of bot-like behavior (likely to have progenitors with intentions to influence).
Our work contributes a substantial and longitudinal online misbehavior dataset to the research community and establishes the consistency of a profile's toxic behavior as a useful factor when exploring misbehavior as potential accessories to influence operations on OSNs. 
\end{abstract}

\section{Introduction}
\label{sec:intro}

Influence operations are organized attempts on Online Social Networks (OSN) to shape people's opinion.
Among strategic tools used by these malefactors are false news, misbehavior against communities based on religion, demographics or sexual orientation, paid trolls and automation (e.g., bots)~\cite{doi:10.1126/sciadv.abb5824}.
The public consensus is that OSNs must take action about malign influence operations, as it is crucial to be able to identify, characterize and predict the profiles instrumented to perform such operations. 

Works such as~\cite{6726818} and \cite{IKEDA201335} have characterized, interpreted and measured Twitter users' misbehavior in specific limited domains and in the user networks, 
while \citet{ribeiro2018like} report differences in content shared by normal and hateful users. 
It is also expected that groups of profiles may work together to create a deeper impact. 
Coordinated efforts of Twitter profiles to spread toxicity or controversies about topics like Bitcoin, etc., were studied by~\citet{pacheco2021uncovering}, who examined networks of coordinated Twitter accounts by analyzing their profile activity and shared media on arbitrary lengths of time.
However, they did not explore  
prolonged involvement of a profile in spreading toxic content and its utility in identifying and characterizing coordination.

In this paper, we seek to identify user profiles consistently pushing toxic content to promote or support influence operations. To this end, we curate the \emph{Twitter Toxic Tweets (3T)} dataset, the largest Twitter dataset to date on this topic, with more than 293M tweets.
This dataset allows us to understand how misbehavior has evolved on Twitter, while its' analysis is foundational in furthering the detection and understanding of Twitter profiles dedicated to spreading a toxic narrative, and their differentiation from other, benign profiles.
{\bf 3T} is seeded with seven smaller public datasets from past works studying online misbehavior on Twitter covering  
multiple themes of online misbehavior: 
hostility, racism, abuse, hatefulness, homophobia, spam and sexism. 
These datasets are balanced in their toxic and non-toxic users. 
A key limitation of the seed datasets is that users are often classified as toxic or not from the content of a single or a handful tweets, which does not allow deeper analysis of the users' ongoing behavior. 
To enable such analysis, we crawl the tweet timeline of each of the users present in the seed datasets. 
Our resulting {\bf 3T} dataset contains 122,255 Twitter profiles and 293,401,161 individual tweets posted between 2007 and 2021.
Human annotations are 
untenable given the size of our dataset.
Hence, we turn to Google's Perspective API models to assign toxicity scores to each tweet, covering the following types of misbehavior: 
\emph{Toxicity, Severe Toxicity, Identity Attack, Inflammatory, Threat, and Insult}. 
To our knowledge, this is the largest dataset annotated with these six Perspective API's models. 

We identify six sets of user profiles who are both posting highly toxic content (based on the median of their tweets scores) and doing so consistently through tweet timeline (based on the Gini index of their tweets scores). 
The selected profiles become the focus of our study. 
We contrast these profiles with sets of random profiles to explore the likelihood they are participants within influence operations furthering hatred and toxicity online. 
We explore the following questions: 
1) What toxic content, URLs, domains and hashtags do they share? What are their high-level topics of interest? How coherent and readable are their tweets? 
2) How homogeneous are these clusters of toxic profiles based on the web resources they share? 
3) Is there a measurable degree of automation within these profiles? 
Armed with this understanding of behavior from potential agents of influence operations, we hope our works inform the creation of improved tools to mitigate their negative impacts on OSNs. 
We demonstrate that our methodology is a useful tool for identifying and understanding toxic influence operations on OSNs. 
Our work provides tools for 
social media moderators to help curtail persistently toxic profiles and to maintain a safe environment for discourse between users. 
This paper makes the following main contributions: 

\begin{itemize}[itemsep=0pt]
    \item {\bf Longitudinal online misbehavior dataset.} We collect and automatically annotate a large longitudinal dataset consisting of 293 million tweets (\S\ref{sec:dataset}). To our knowledge the largest published dataset on online misbehavior. 
    Upon publication, we plan to release our enriched dataset for future research. 
   
    \item {\bf Identification of consistently misbehaving profiles.}
    Using Gini index and toxicity scores, we propose a novel approach to identify profiles that are consistently generating online toxic content, and demonstrate that this is an effective tool for identifying profiles likely involved in toxic influence operations 
    (\S\ref{sec:focus-users}). 
    
    \item {\bf Characterization of consistently misbehaving profiles.}
    We characterize profiles for casual vs. consistent misbehavior (\S\ref{sec:less diverse content}). 
    We observe that consistently toxic profiles are specific and cohesive in types of shared web content. 
    Hashtags shared by such profiles are coherent, but toxic and malignant in nature, which sets them apart from profiles occasionally involved in online misbehavior.
    Such profiles persistently discuss toxic and sensitive topics about war zone, ethnicity and religion. 
    \item {\bf Analyzing homogeneous temporal misbehavior of profiles across categories of misbehavior.}
    Our focus profiles across all categories share comparable number of similar domains.
    We identify consistently misbehaving profiles sharing interests via embedding similar hashtags (\S\ref{sec:homogeneity}). 
    Our study signifies that consistently toxic users maintain a very discrete tweeting pattern, which extends to using specific hours of the day and specific days of the week.
    We reveal that it is highly likely that such misbehaving profiles are automated accounts instrumented for spreading misbehavior (\S\ref{sec:time-analysis}). 
\end{itemize}

\section{Dataset Methodology and Characterization}
\label{sec:dataset}

\begin{figure}[t]
  \centering
  \includegraphics[width = 1.0\columnwidth]{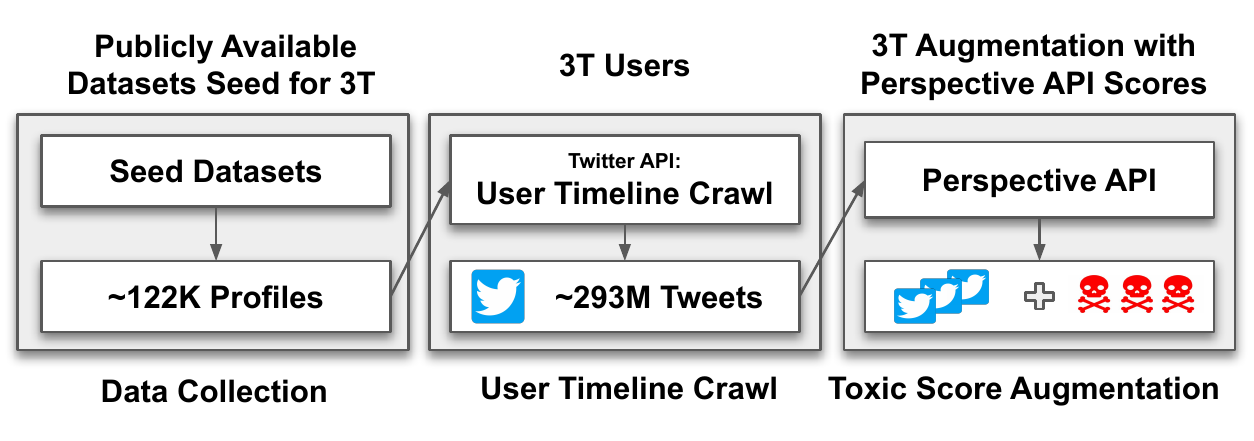}
  \vspace{-4mm}
  \caption{Twitter Toxic Tweet (3T) dataset collection and augmentation pipeline.
  }
  \label{fig:data_collection_flowchart}
  \vspace{-4mm}
\end{figure}
In this section, we detail our methodology for data collection and augmentation (overview shown in Figure~\ref{fig:data_collection_flowchart}).
Section~\ref{sec:seed-datasets} states our seed datasets, and Section~\ref{sec:user-timeline-crawl} provides details on the crawling of user timelines.
Section~\ref{sec:3T-augmentation} details our efforts to augment the crawled data with toxicity scores by querying the Google Perspective API, and finally, Section~\ref{sec:3T-characterization} provides a characterization of the augmented dataset.

\begin{table*}[!th]
\centering
\resizebox{1.0\textwidth}{!}{
\begin{tabular}{c|r|r|r|l|l||r|r}
\toprule
\multicolumn{6}{c||}{\bf Seed Dataset} & \multicolumn{2}{c}{\bf Crawled Dataset}\\\midrule

Dataset & TIDs &  UIDs & RUIDs & Labels or Keywords & Collection Method & CUIDs & Tweets Retrieved \\ \midrule

\citet{gomez2019exploring} & 150,000 & - & 895 & sexist, racist, homophobic,  religion, other hate, no hate    &   Amazon Mechanical Turk     & 841   & 1,550,654   \\ \hline 
\citet{kaggle:metoomovement} & 807,174 & - & 19,859 &  keyword:MeTooMovement    &  Twitter Streaming API   & 15,085          & 28,684,131  \\ \hline
\citet{ribeiro2018like} & - & 100,386 & 100,386 & hateful, not hateful    &  CrowdFlower (appen)  & 55,125            & 115,429,956\\ \hline
\citet{founta2018large} & -  & 98,377 & 98,337 & normal, abusive, spam, hateful    & CrowdFlower (appen)          & 35,125 & 75,151,305 \\ \hline
\citet{jha-mamidi-2017-compliment}   & 10,583 & - & 324 & benevolent, hostile, other    & SVM (TF-IDF)        & 310 & 574,353   \\ \hline
\citet{waseem-hovy-2016-hateful} & 16,907 & - & 891 & sexist, racist, neither    &  CrowdFlower (appen)  & 154    & 253,326   \\ \hline
\citet{waseem-2016-racist}  & 6,909 7 & - & 870 & sexist, racist, both, neither    &  CrowdFlower (appen)   & 23 & 1,900,539  \\ \midrule
\textbf{3T} & & & & {\bf Perspective Scores (Continuous)}     &   & \textbf{122,255} &  \textbf{293,401,161}\\ \bottomrule
\end{tabular}
}
\caption{
Overview of 7 datasets used as seed with a collection of User IDs (UIDs) or Tweet IDs (TIDs), whatever was made publicly available.
TIDs were used to recover Users IDs (RUIDs).
Crawled UIDs (CUIDs) are the user profiles successfully crawled.
CUIDs can be smaller than RUIDs if said RUID Twitter profiles were not found.}
\label{tab:dataset_info}
\vspace{-4mm}
\end{table*}

\subsection{Data collection}
\label{sec:seed-datasets}

We curated 7 publicly available datasets from past studies, as our base dataset.
All these studies used human annotation, existing ML models or random sampling on streaming API to label users as toxic or not (based on a single tweet).
A summary of the selected datasets, details of their size and labels can be found in Table~\ref{tab:dataset_info}.

\subsection{User timeline crawl}
\label{sec:user-timeline-crawl}
As per Twitter Terms and Conditions, Twitter User IDs (UIDs) and tweet content cannot be publicly shared. 
Consequently our seed datasets contained Tweet IDs (TIDs) and their respective annotation.
Therefore, the first step was to query Twitter's API~\cite{twitterAPI} to recover the UID responsible for each TID.
Next, we queried Twitter API with these UIDs to retrieve each profile's historical tweets.
The Twitter API allows the retrieval of the 3,200 most recent tweets from a profile, allowing us to study the historical record of each user and their evolving toxic behavior.
We were unable to retrieve tweets from banned, deleted or private profiles.
From the retrieved tweets (JSON files), we extracted relevant details such as the text, date of creation, hashtags, URLs and domains shared within tweets.
For this study, we only consider English tweets.
While the investigation of other languages can offer additional insights, we were constrained by Perspective API's range of supported languages at that time.
Additionally, it was unclear if scores between languages are calibrated and directly comparable.

\subsection{Dataset augmentation with Perspective API}
\label{sec:3T-augmentation}

In the aforementioned seed datasets, one tweet was annotated or labeled per user.
However, it is unrealistic to assume that this single-tweet label can be propagated on all the tweets of said user.
Thus, to obtain a measure of misbehavior across all tweets of each user, we employed Google's Perspective API~\cite{perspective}.
Perspective API provides multiple Convolutional Neural Networks (CNNs)-based models trained with GloVe word embedding~\cite{pennington-etal-2014-glove} for the evaluation of misconduct in any submitted text.
This API offers 16 ML models that can provide a probabilistic score from 0 to 1 on the given text having an intensity on a specific dimension such as Toxicity, Threat, Inflammatory, etc.
We focused on receiving scores from this API on the following six dimensions, defined as:
\squishlist
    \item \emph{Toxicity}: Rude, disrespectful, or unreasonable comments, likely to make people leave a discussion.
    \item \emph{Severe Toxicity}: Comments very likely to make users leave a discussion or give up sharing their perspective.
    \item \emph{Identity Attack}: Negative or hateful comments targeting someone because of their identity, ethnicity, sexual orientation and such.
    \item \emph{Inflammatory}: Intended to provoke or inflame.
    \item \emph{Insult}: Insulting or negative comments towards a person or a group of people.
    \item \emph{Threat}: Intentions to inflict pain, injury, or violence against an individual or group.
\squishend
The Perspective API provides multiple other experimental dimensions which we do not use here.
We polled all 293M English tweets for a score from each of these models.
Thus, each tweet in our dataset has these six scores.

\begin{figure*}[t]
    \centering
    \begin{subfigure}[t]{0.245\linewidth}
            \includegraphics[width=\textwidth]{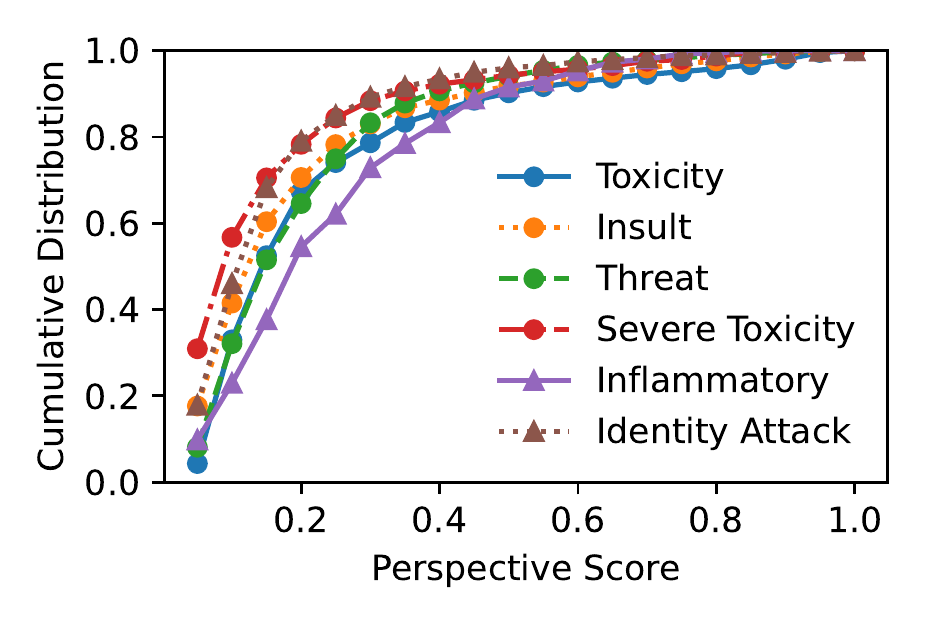}
            \caption{\small Perspective API scores}
            \label{fig:all_tweet_scores}
    \end{subfigure}
    \hfill
    \begin{subfigure}[t]{0.245\linewidth}
            \includegraphics[width=\textwidth]{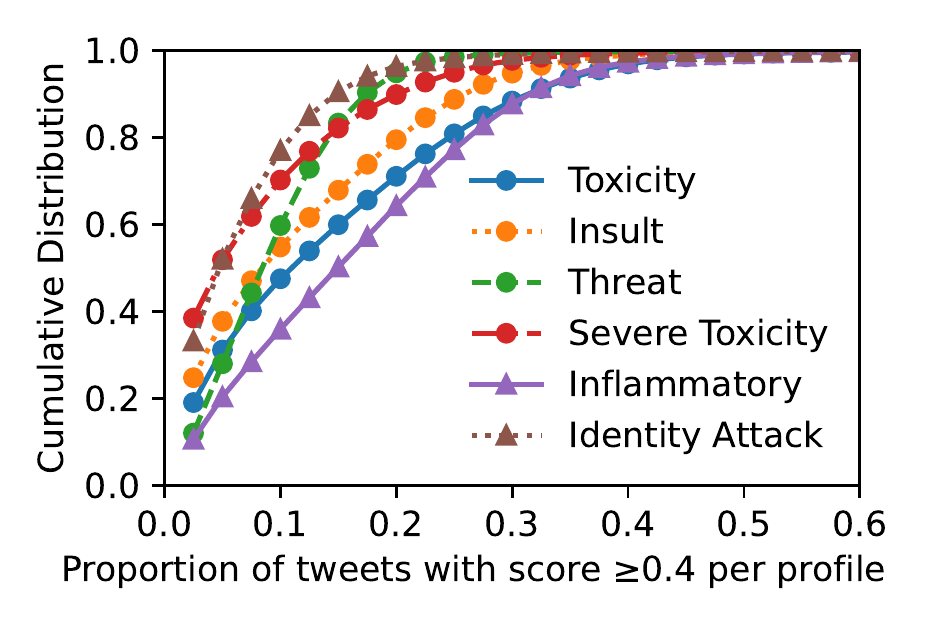}
            \caption{\small Ratio of tweets for score$\geq$0.4}
            \label{fig:toxpercent_user_tweet_st}
    \end{subfigure}
    \hfill
    \begin{subfigure}[t]{0.245\linewidth}
            \includegraphics[width=\textwidth]{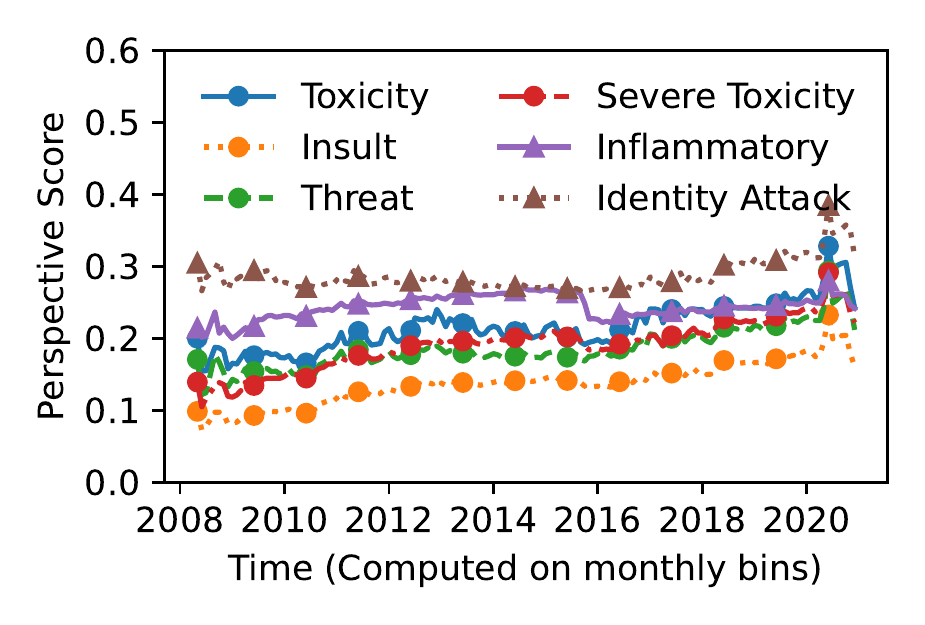}
            \caption{\small Median score per month}
            \label{fig:med_by_month}
    \end{subfigure}
    \hfill
    \begin{subfigure}[t]{0.24\linewidth}
            \includegraphics[width=\textwidth]{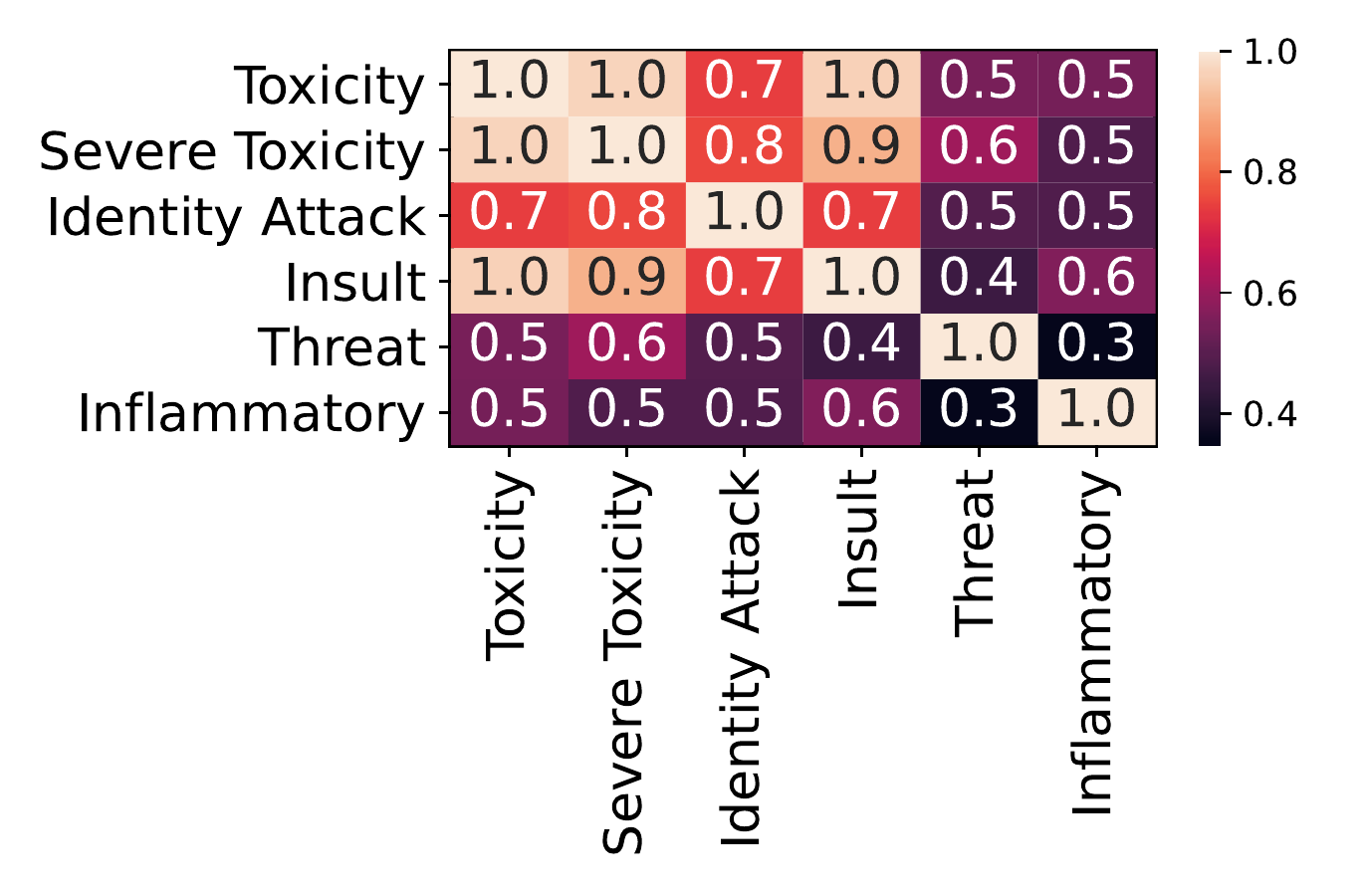}
            \caption{\small Correlation between scores}
            \label{fig:cor_matrix}
    \end{subfigure}
    
    \caption{(\protect\subref{fig:all_tweet_scores}) Cumulative Distribution Function (CDF) for each Perspective API score, across all tweets; (\protect\subref{fig:toxpercent_user_tweet_st}) CDF of profile proportions of misbehaving tweets (tweets that equal or exceed a Perspective API score of 0.4) for all profiles; (\protect\subref{fig:med_by_month}) Median Perspective API score of all tweets per month, broken down in categories of misbehavior; (\protect\subref{fig:cor_matrix}) Pearson pairwise correlation matrix across all Perspective API scores, computed on all tweet scores of 3T.}
    \label{fig:characterize} 
    \vspace{-4mm}
\end{figure*}

\subsection{Characterization of augmented 3T}
\label{sec:3T-characterization}

To better understand the composition of our curated dataset, we first inspect the Cumulative Distribution Function (CDF) of each Perspective score, across all tweets through time (Figure~\ref{fig:all_tweet_scores}).
We observe that the median score of a tweet for any of the six dimensions varies in the range 0.1 -- 0.2.
Also, a steady rise in the curve in the low ranges of scores indicates that a majority of tweets do not strongly exhibit any specific form of misbehavior (80\% of tweets have scores less than 0.4).
Also, the strongest signal for misbehavior is in the dimension of Inflammatory content.
A tail is also observed of tweets acting as exception to the rule, propagating what is perceived as a large amount of misbehavior (score $\rightarrow$ 1.0).

Following other studies (e.g.,~\citet{elsherief2018peer}), we attempt to binarize the tweets are misbehaving or not, by applying a threshold of 0.4 (as an example) on each score (we discuss later in Section~\ref{sec:focus-users} the use of 0.4 as threshold).
Then, we compute and plot in Figure~\ref{fig:toxpercent_user_tweet_st} the proportion of misbehaving tweets each user has posted.
We find that 80\% of all users have a maximum of 30\% of tweets (or less, depending on misbehavior category) considered as misbehaving.
Still, there is a tail of strongly misbehaving users with a high percentage of tweets meeting this condition of misbehavior.

Furthermore, to inspect how misbehavior has changed through time across Twitter as a platform (represented by our sample), we bin tweets per month, and compute the median score across all user tweets created during each monthly bin.
These median scores are presented in Figure~\ref{fig:med_by_month} and demonstrate an increase in the level of misbehavior through time, especially in the years 2016-2020.
In particular, when computing a linear regression model for each toxic behavior through time (Table~\ref{tab:trend_misbehave_regression}), we find that all six categories fit well such a model (with highest p-value of an F-test being $7\times10^{-6}$), and positive slopes ranging from 0.092 to 0.268. These trends may be potential indicators for expanded influence operations in recent years.

\textbf{Takeaway 1:}
Toxic behavior on Twitter has been increasing through the last 15 years across six dimensions of misbehavior in tweets' texts.

\begin{table}[t]
\resizebox{\columnwidth}{!}{
\begin{tabular}{l|cccccc}
\toprule
                                & Tox. & Insult & Threat & Sev. Tox. & Inflam. & ID Attk. \\ \midrule
Coeff     & 0.169    & 0.092  & 0.144  & 0.139            & 0.234        & 0.268            \\ \hline
$R^2$     & 0.739    & 0.870  & 0.765  & 0.850            & 0.127        & 0.273            \\ \hline
P(F-stat) & $3*10^{-45}$    & $7*10^{-68}$  & $1*10^{-48}$  & $3*10^{-63}$           & $7*10^{-6}$        & $6*10^{-12}$            \\ \bottomrule
\end{tabular}
}
\caption{Linear regression (Ordinary Least Squares) of median Perspective score of all collected tweets by month (as plotted in Figure~\ref{fig:med_by_month}).}
\label{tab:trend_misbehave_regression}
\vspace{-4mm}
\end{table}

\section{Ethical Considerations}

The research presented in this paper is non-commercial, in line with Twitter’s Terms and Conditions for research purposes.
We used standard Twitter API to collected publicly available data on Twitter, from tweets of public user profiles.
We acknowledge the responsibility of security and privacy which comes with the data collected.
During the storing and processing of the data, Twitter users were referred to only by their UIDs.
In all of our experiments, any result produced and shown cannot be used to re-identify, or track said users, as no user profiles are specifically named.

During our experiments, we follow ethical guidelines outlined in~\citet{rivers2014ethical}.
Given our experimentation on human-produced data, we obtained formal ethics committee approval from our institution's IRB.
Our data will not be shared with any third-party for commercial purposes.\footnote{Macquarie University IRB Project Reference: 35379, Project ID: 10008, Granted: 27/11/2021}

Our work is first to release scores for six Perspective API models for 122K profiles and 293M tweets to help facilitate the research in combating the online misbehavior in six dimensions on Twitter.
The released {\bf 3T} dataset will be a collection of TIDs and six Perspective API scores per TID: Toxicity, Severe Toxicity, Identity Attack, Inflammatory, Insult and Threat.
Sharing of TIDs and scores are inline with Twitter and Perspective API's Terms and Conditions.
We expect this work to contribute to a broader understanding of online misbehavior, by identifying consistently misbehaving Twitter user profiles which contribute disproportionately to online toxic content, as well as profiles that may be unintentionally marked as such due to errors in API scores.

\section{Identification of Focus Profiles}
\label{sec:focus-users}

\begin{figure}[t]
    \centering
    \includegraphics[width=0.80\linewidth]{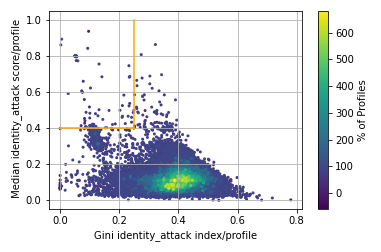}
    \vspace{-0.5mm}
    \label{fig:toxic-gini-med-IDATTK}
    \caption{Focus profiles for Identity Attack scores (bounded by the orange box), based on median and Gini coefficient of Identity Attack scores of all 3T profiles.}
    \label{fig:toxic-IDATTK}
    \vspace{-5mm}
\end{figure}

Our goal is to identify and study profiles serving in influence operations, that consistently post tweets high in the six categories of misbehavior: Toxicity, Severe Toxicity, Identity Attack, Inflammatory, Insult and Threat.
For each type, we identify sets of consistently toxic profiles based on the median score and Gini index of each profile's scores (\S\ref{sec:clustering}).
We refer to these profiles as `focus profiles'. 
We then perform a first-pass filtering to remove obscene content (\S\ref{sec:remove_obscene}).

\subsection{Twitter profile clustering}
\label{sec:clustering}

To identify focus profiles, we cluster the available profiles with respect to two main axes: overall high level of misbehavior (represented by median Perspective scores), and overall low variability in the toxicity of their posts. To measure a user's variability in toxicity, we use the Gini coefficient of their tweets' Perspective scores. The \emph{\textbf{Gini coefficient}} was originally intended as a measure of the concentration of wealth~\cite{gini1912variabilita}, but can equally be used to identify evenly distributed values such as toxicity, in our case. A consistent set of (low or high) scores produces a value closer to 0, whereas a wide range of (both low and high) scores produces a Gini coefficient value closer to 1. We drop profiles with less then 10 tweets, to have enough activity per user and reliably compute the two metrics.
\begin{table}[t]
\resizebox{\columnwidth}{!}{
\begin{tabular}{l|cc}
\toprule
{\bf Perspective Model} & {\bf \#;\% Focus Profiles (Tweets)} & {\bf \#;\% Random Profiles (Tweets)} \\
\midrule
Inflammatory     & 161; 0.13\% (294,264 tweets)      & 161; 0.13\% (362,381 tweets)\\
Toxicity         & 78; 0.06\% (147,332 tweets)       & 78; 0.06\% (167,026 tweets)\\
Identity Attack  & 65; 0.05\% (115,805 tweets)       & 65; 0.05\% (158,645 tweets)\\
Insult           & 52; 0.04\% (114,404 tweets)       & 52; 0.04\% (142,192 tweets)\\
Severe Toxicity  & 46; 0.03\% (92,518 tweets)        & 46; 0.03\% (105,007 tweets)\\
Threat           & 34; 0.02\% (54,935 tweets)        & 34; 0.02\% (78,640 tweets)\\
\bottomrule
\end{tabular}
}
\caption{Number of focus profiles per Perspective API dimension. An equal number of random profiles are selected for comparison.
\vspace{-3mm}
}
\label{tab:toxic-random-groups}
\vspace{-5mm}
\end{table}

We set a lower threshold on the median (0.4) and upper threshold on Gini coefficient values (0.25) to capture extreme and consistent toxic behavior. 
This approach is similar to that of~\citet{elsherief2018peer} where they choose 0.8 for Toxicity and 0.5 for Attack on Commenter scores (one of the Perspective API dimensions) to identify very toxic tweets in their corpus. We assume a conservative threshold of Gini in the range of 0.0 -- 0.25 aiming to capture highly consistent toxic behavior expected from participants in influence operations. The process is illustrated in the scatter-plot of Figure~\ref{fig:toxic-IDATTK} for Identity Attack scores.
Each dot represents a Twitter profile in 3T and the yellow box indicates the aforementioned thresholds. 
The profiles (dots) falling within the yellow box are our focus profiles for the Identity Attack dimension.
This step is repeated for all six Perspective scores (for the plots for Inflammatory,Insult and Threat scores please refer to Appendix Sec. ~\ref{all focus groups}). 
The resulting six clusters of focus profiles are summarized in Table~\ref{tab:toxic-random-groups}.
For the rest of the paper, \textbf{\emph{Focus Profiles}} are referenced based on the type of misbehavior they represent.
In addition, for every set of focus profiles, we also select a random set of profiles from the 3T data, equal in number with each set of focus profiles.
We refer to these as \textbf{\emph{Random profiles}}, per type of misbehavior. As seen in Table~\ref{tab:toxic-random-groups}, the selected thresholds of values for identifying focus profiles lead to a small number of such profiles in all clusters, compared to the size of the 3T dataset.
This can be attributed to the following reasons:
a) to start with, the number of toxic tweets on Twitter is expected to be generally small ($\sim$8\% was reported by~\citet{founta2018large});
b) then, extremely toxic profiles do not last long on Twitter: they get reported for misconduct violation and are banned fairly quickly~\cite{10.1145/3479525}; 
c) our selected thresholds are conservative (min median=0.4 and max Gini=0.25) and applied in a combined fashion, as we aim to identify extreme cases of both highly toxic profiles, who are also consistent in their toxicity (not just sporadically toxic).
\subsection{Removal of consistently obscene focus profiles}
\label{sec:remove_obscene}

In this work, we go beyond past studies that focus on typical Perspective scores such as (Severe) Toxicity~\cite{elsherief2018peer,hosseini2017deceiving,jain2018adversarial}, and study Identity Attack, Inflammatory, Insult and Threat to identify profiles exhibiting diverse type of misbehavior.
Thus, first, we compute a pairwise Pearson correlation across the scores of tweets between the six dimensions and plot their potential signal similarities in Figure~\ref{fig:cor_matrix}.
It becomes clear that Toxicity and Severe Toxicity are highly correlated with each other, and with the rest of the dimensions.

Then, we investigate the type of content shared by the focus profiles based on these scores.
In particular, we count all the hashtags shared by these profiles and investigate the top 50 most frequently shared hashtags in each group. 
In all profile selections, 35\%-42\% of hashtags were obscene in nature.
For example, 31 out of the 50 most shared hashtags in the focus Identity Attack profiles were obscene, e.g. `xxx', `adult', `asian', `nsfw', `beardedmen', and `beards', and 7\% of profiles had shared these hashtags in more than 80-85\% of their tweets.
Indeed, pornographic content is shared liberally on Twitter~\cite{PewResearchCenter2018}.
However, we are not interested in characterizing sexual obscenity on Twitter, instead to find potential participants of influence campaigns.
We note that in the focus Toxicity and Severe Toxicity focus selections, 80-85\% of profiles are involved in tweeting obscene hashtags, in contrast to the other four sets of focus profiles, whereby 6-9\% profiles are responsible for a majority of obscene hashtags.
Thus, like~\cite{gomez2019exploring}, we drop such profiles, after manual inspection of their hashtags.
Overall, given the correlation result on Toxicity and Severe Toxicity, and the highly obscene profiles included in them, we decided to drop these two dimensions, and focus on the other four: Identity Attack, Insult, Inflammatory, and Threat.
We also drop the 6-9\% of profiles in these dimensions whose shared obscene hashtags exceed 80\%.

\section{Content Diversity of Focus Profiles}
\label{sec:less diverse content}

The nature of the text in a profile's tweets, as well as auxiliary content included in the tweets and in the profile, such as URLs and hashtags, can represent the profile's focus and interests. If their interests align under influence operations, shared content of multiple profiles should follow suit.
To extract these interests, we perform a longitudinal analysis of all tweets per focus profile, and observe the nature of shared content, over the four types of misbehavior: Identity Attack, Insult, Inflammatory, and Threat.
In particular, we perform analysis on URLs (\S\ref{sec:url-diversity}) and hashtags (\S\ref{sec:hashtag-diversity}) shared, the topics of tweets (\S\ref{sec:topic-diversity}) and their degree of readability (\S\ref{sec:readability}).

\subsection{How diverse are the URLs shared?}
\label{sec:url-diversity}

For URL analysis, we first detect all URLs from focus profiles per misbehaving dimension (Identity Attack, Inflammatory, Insult and Threat) and the corresponding sets of random profiles.
Then, we extract second level domains (SLD) from all detected URLs, resulting in 319,082
SLDs.
A SLD is the part of the domain that is located right before a Top Level Domain (TLD).
For example, in {\tt{www.example.com}} the SLD is {\tt{example.com}} and the TLD is {\tt{com}}.
Beyond this point when we mention Domain, we refer to the SLD.
Next, we classify these domains with the \textbf{\emph{FortiGuard}} classification service~\cite{Fortiguard}.
FortiGuard uses link crawlers, customer logs and machine learning to categorize websites~\cite{Triplet20}. 
Using FortiGuard, we successfully categorize 312,702 (98\%) domains; the remaining 2\% (6,380) corresponds to 32,419 (0.28\%) of the web pages distributed by user tweets.

\begin{table}[t]
\centering
\small
\tabcolsep=0.05cm
\scalebox{0.75} {
\begin{tabular}{l | r r | r r  | r r  | r r  }
\toprule

           & \multicolumn{2}{c}{\bf Identity Attack} & \multicolumn{2}{c}{\bf Inflammatory}  & \multicolumn{2}{c}{\bf Insult} & \multicolumn{2}{c}{\bf Threat} \\
            \cline{2-9}

		&	Focus	&	Random	&	Focus	&	Random	&	Focus	&	Random	&	Focus	&	Random	\\
	\midrule

\# URLs		&	62,338	&	42,988	&	55,611	&	51,954	&	71,897	&	36,720	&	3,116	&	3,408		\\
\# Uniq. URLs		&	57,762	&	42,962	&	55,607	&	51,896	&	67,320	&	36,678	&	3,116	&	3,408	\\
Avg. Uniq. URLs		&	2,406.8	&	1,718.5	&	1,425.8	&	1,235.6	&	2,321.4	&	1,222.6		&	779	&	681.6	\\
\hline
\# Domains		&	62,338	&	42,988	&	55,611	&	51,954	&	71,897	&	36,720		&	3,116	&	3,408	\\
\# Uniq. Domains	&	51	&	1,100	&	585	&	977	&	136	&	1,027		&	33	&	68	\\
Avg. Domains		&	2.1	&	44.0	&	15.0	&	23.3	&	4.7	&	34.2	&	8.2	&	13.6		\\\hline

\# Domains Cat.		&	20	&	49	&	43	&	48	&	29	&	45	&	16	&	21	\\

\bottomrule
\end{tabular}
}
\caption{Breakdown of URLs, domains and categories of domains for focus vs. random profiles, across four categories of misbehavior.
}
\vspace{-0.2cm}
\label{tab:domain_url_cat_stats}
\end{table}

\begin{figure}[t]
\centering
\includegraphics[width=0.70\columnwidth]{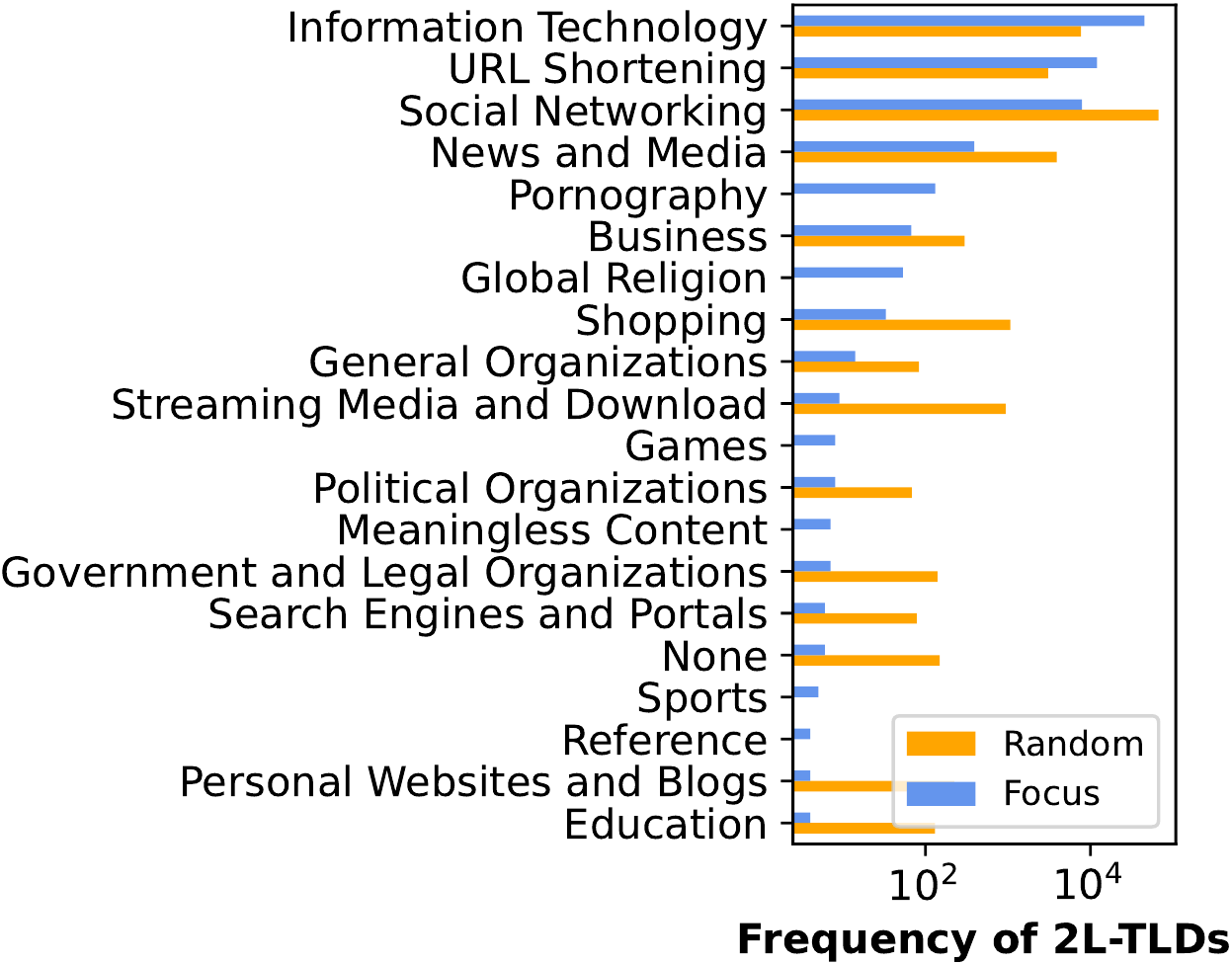}\label{fig:median:IDENTITY_ATTACK_2-category}
\caption{Top 20 domain categories in focus Identity Attack profiles. Here, 2L-TLDs refer to second level domains (SLDs). 
``None'' refers to unrated websites whose domain category is unknown to FortiGuard.}
\label{fig:bar_url_puser:gini_median-category}
\vspace{-0.2cm}
\end{figure}

Table~\ref{tab:domain_url_cat_stats} provides a breakdown of total and unique number of URLs, domains and domain categories for focus vs. random profiles.
We note that not every focus profile shared URLs in their tweets, so the average of unique URLs per profile is computed using only the profiles which shared URL(s).
We observe that all focus profiles shared a larger number of total URLs, and total domains than random profiles.
Indeed, when we look at the unique URLs and domains, focus profiles present a different picture: they have shared a larger number of unique URLs, but a much smaller number of unique domains, than random profiles.
In particular, the focus Insult profiles shared the highest number of total (71,897) and unique (67,320) URLs, and almost double that of random profiles.

Perhaps expectantly, random profiles shared URLs that are fairly unique (since their total and unique number of URLs is almost the same).
When we look into unique domains, we observe a different picture: focus profiles share a much smaller (in some cases orders of magnitude less) number of unique domains than random.
For example, focus Identity Attack profiles referenced only 52 unique domains, compared to 1,100 from random profiles, even though the two sets have same number of profiles and same order of magnitude total number of URLs.
We also observe similar results in focus Insult profiles which shared only 136 unique domains from 71,897 URLs, compared to 1,027 unique domains from 36,720 URLs.

These results are also reflected in the average number of unique URLs and domains per profile: for all focus profile types, the averages are smaller than random profiles, demonstrating that focus profiles are, on average, sharing less diverse set of URLs and domains.
Also, looking into domain categories extracted from Fortiguard, all sets of focus profiles have smaller number of categories than random sets of profiles.

We look into this further by plotting the top 20 categories of domains out of 89 different categories found in focus Identity Attack profiles in Figure~\ref{fig:bar_url_puser:gini_median-category} (Similar plots for other categories of misbehavior can be found in Appendix Sec.~\ref{url analysis}).
We observe that, focus identity attack profiles share different types of domains, and with different intensity, than random profiles.
Clear trends can be seen, with Information Technology (e.g., {\tt marinsoftware.com} and {\tt nec.com}), URL Shortening (e.g., {\tt fb.me}, {\tt tinyurl.com} and {\tt myburbank.com}), Social Networking (e.g., {\tt facebooklive.com}), and News \& Media (e.g., {\tt unfoxnews.com}).
Also, random profiles share a broader spectrum of web resources, from a more uniform distribution of categories (social networking, business, entertainment, shopping, streaming media, politics, etc.)
The same trends are present in the other types of misbehaving profiles, with the top 3 categories of Inflammatory and Insult including Information Technology, URL Shortening, and Social Networking domains.
The exception is Threat, where top 3 contains Social Networking, News \& Media and Streaming Media \& Download.
Finally, pornographic content dominates focus profiles of Identity Attack and Insult, and is present in all toxic clusters, with 0.2\%, 0.02\%, 0.17\%, and 0.12\% in Identity Attack, Inflammatory, Insult, and Threat, respectively.
These results are in line with findings of~\citet{PewResearchCenter2018}.

Finally, Figure~\ref{fig:url_domain_puser:gini_median} shows the CDF of number of unique URLs and domains per focus Identity Attack vs. random profile. For results based on other categories of misbehavior please refer Appendix ~\ref{sec:no.url/domains}.

We observe that 95\% of focus profiles post at most 100 unique domains, while 20\% of the random group of users share at least 100 unique domains in their tweets.
This suggests a tendency of such Identity Attack profiles to share a narrower set of external web content than random profiles.

\textbf{Takeaway 2:}
Focus profiles fetch very specific and cohesive in type web content, originating from a larger set of URLs, compared to random profiles.
Random profiles share a wider range of domains from a smaller set of URLs, pointing to more diverse web resources included in their tweets.

\subsection{How diverse are the hashtags shared?} 
\label{sec:hashtag-diversity}

Adding hashtags to tweets is a popular and easy way for users to convey a message to an interested audience, and to have a voice in intended communities.
In order to compare the tendency of sharing hashtags from focus and random profiles, we extracted and compared the total and unique number of hashtags in focus and random profiles.

\begin{figure*}[!htb]
\centering
    \subfloat[{\small URLs \& domains in tweets}]{
    \includegraphics[width=0.23\textwidth]{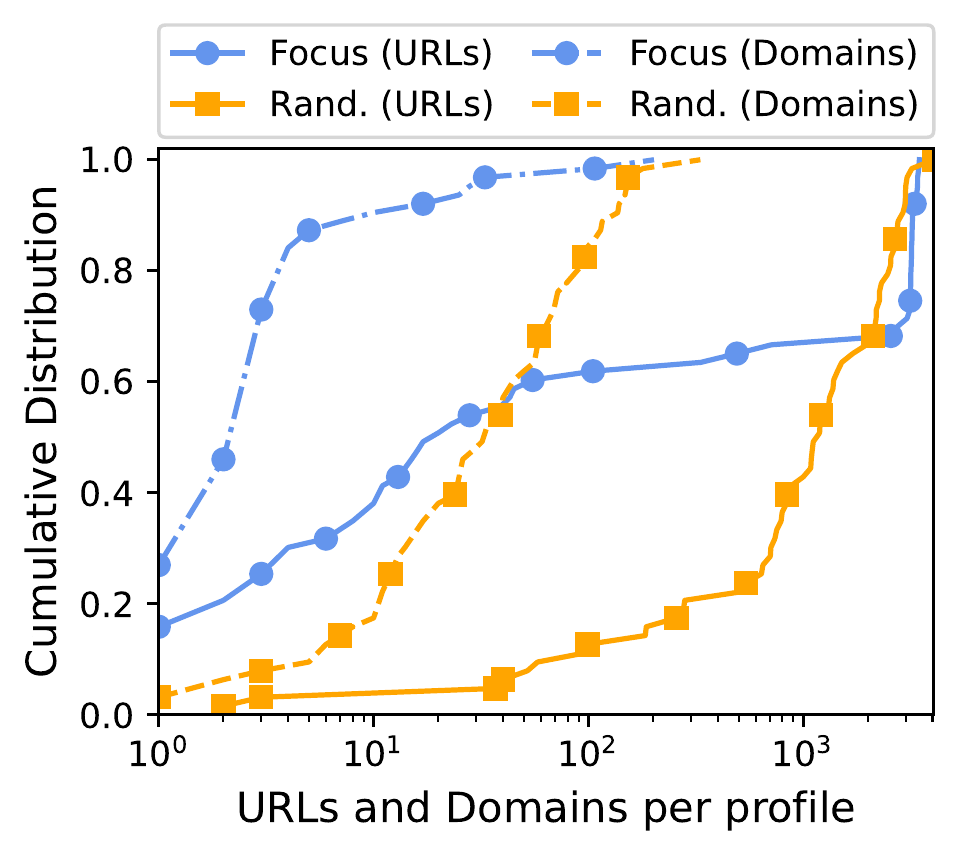}
    \label{fig:url_domain_puser:gini_median}
    }\hfill
    \subfloat[{\small Hashtags in tweets}]{
    \includegraphics[width=0.23\textwidth]{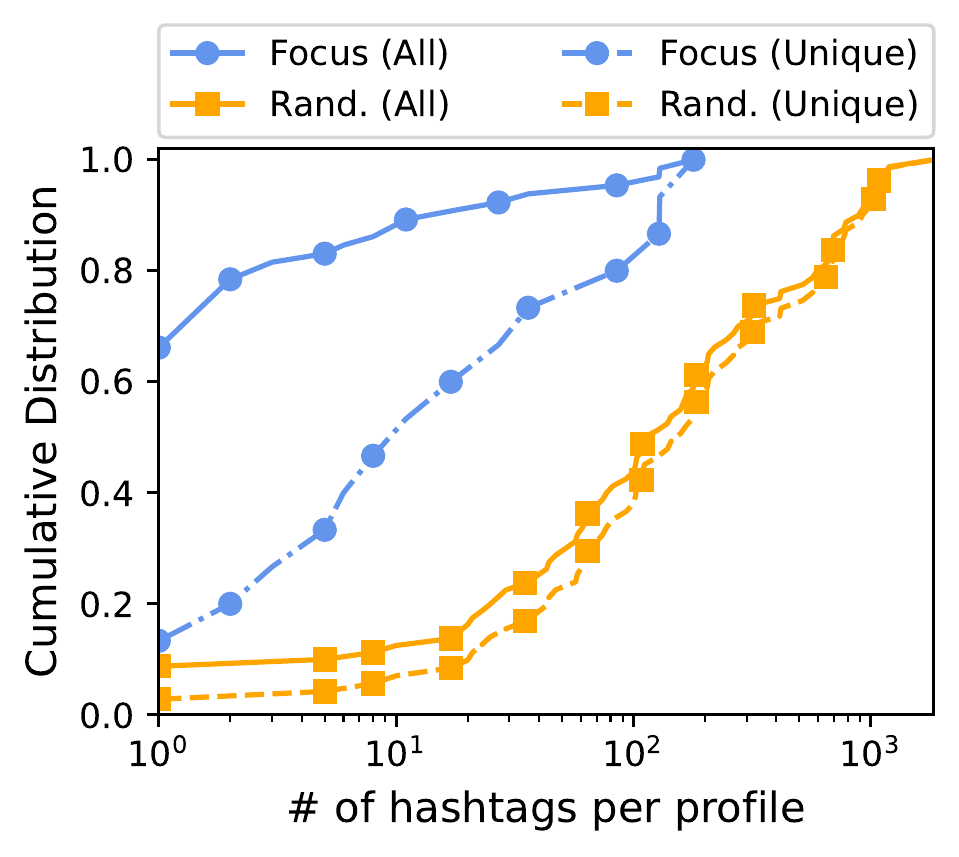}\label{fig:url_htags_similarity:gini_median-h}
    }\hfill
    \subfloat[{\small Similarity of domains}]{ 
    \includegraphics[width=0.23\textwidth]{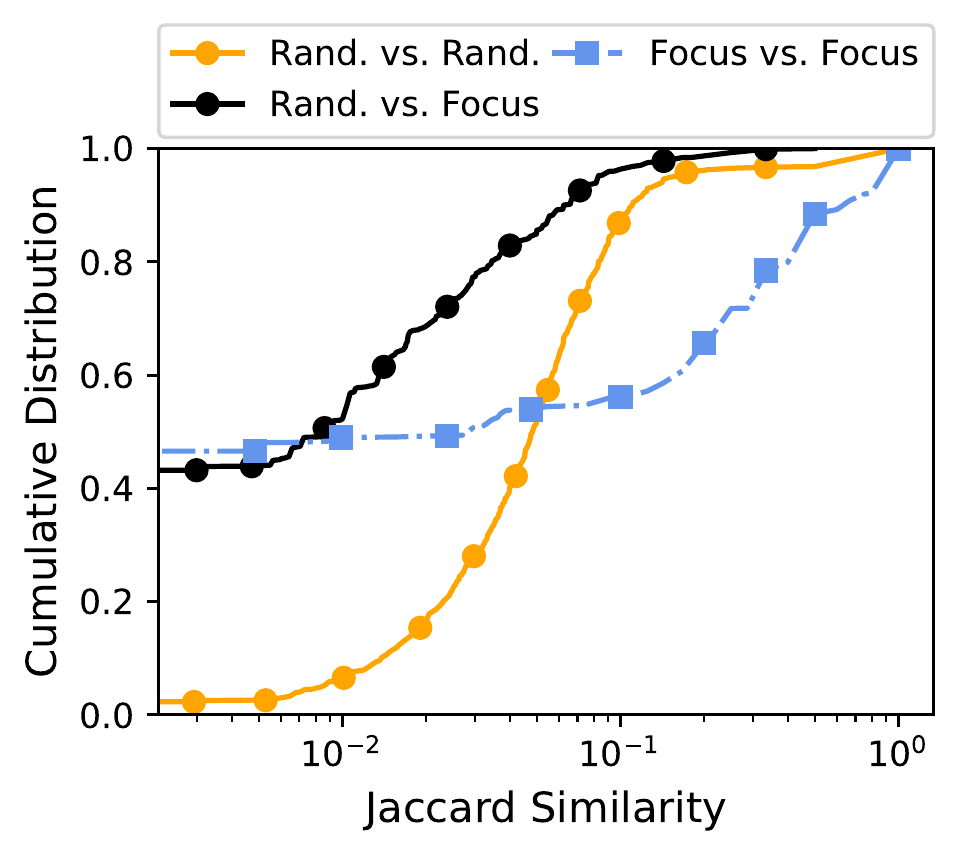}\label{fig:url_domain_puser:gini_median-dom-simm}
    }\hfill
    \subfloat[{\small Similarity of hashtags}]{ 
    \includegraphics[width=0.23\textwidth]{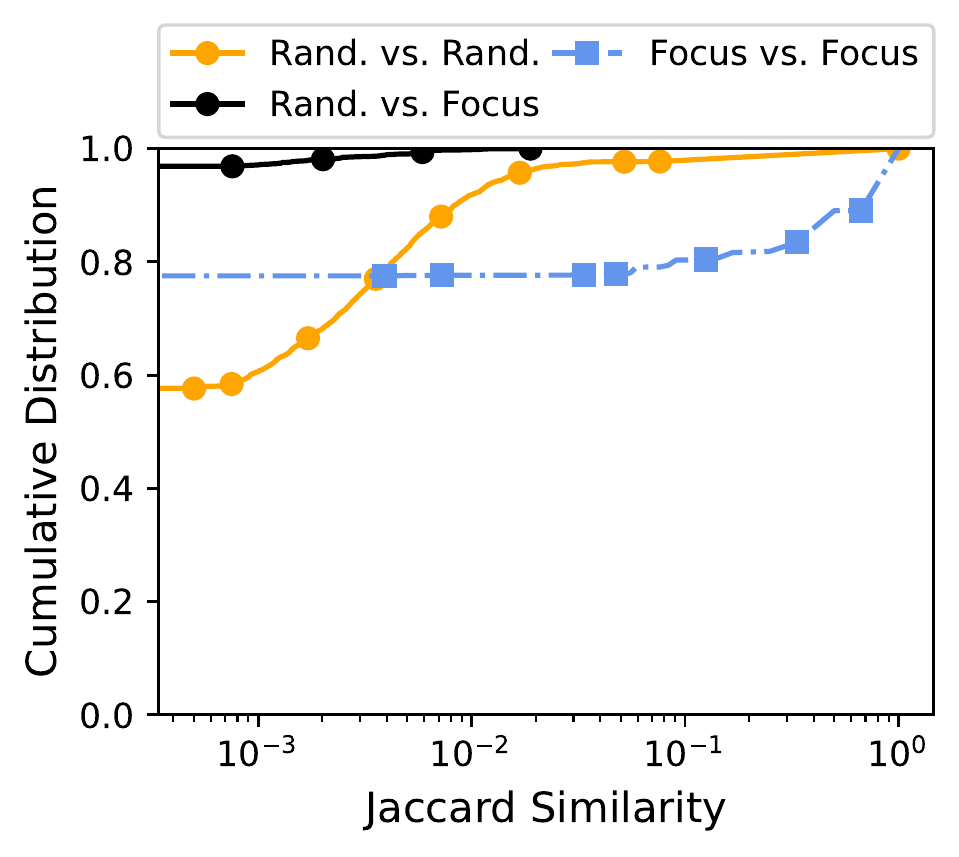}\label{fig:url_htags_similarity:gini_median-simm}
    }
\vspace{-0.1cm}
\caption{CDFs of number and similarity of hashtags and URLs computed for Identity Attack.}
\vspace{-0.2cm}
\end{figure*}

Figure~\ref{fig:url_htags_similarity:gini_median-h} shows the CDF of these counts per profile, in focus Identity Attack, and random profiles. More results based on focus Inflammatory, Insult and Threat and their random sets of profiles can be found in Appendix~\ref{sec:no.of hashtags}.

Around 70\% of the focus profiles do not use any hashtags in their tweets, whereas there is only $<$1\% of random profiles with no hashtags, suggesting that hashtags are generously used by all sets of random profiles.

On the focus profiles that do use hashtags, 90\% of them use at most 10 hashtags, in contrast to 50\% of random profiles that use at least 100 hashtags, demonstrating the diverse interests covered by random profiles.

Table~\ref{tab:thags} shows that focus Identity Attack profiles share least number of total 701, and unique 612, hashtags.
On the other hand, the focus Inflammatory profiles share the highest number of total (5,299) and unique (4,008) hashtags.

Overall, the four types of focus profiles share a considerably smaller number of hashtags than random profiles, and choose to engage specific and very few communities through hashtags.

Diving into the hashtags posted by these focus profiles (Table~\ref{tab:top5_htags_gini_median}), they are strikingly different in nature to the ones from random profiles.
Focus Identity Attack profiles share hashtags about warfare and conflict.
The most shared hashtag \emph{\#TreCru} is about Treasure Cruise, an action role play combat game, whereas other hashtags include countries under attack or in war situation like \emph{\#Syria} and \emph{\#BDS} (i.e., \emph{Boycott, Divestment, Sanctions}).
This is a Palestinian-led movement for freedom and equality.
Random profiles, on the other hand, share hashtags which indicate major happenings such as \emph{\#Covid, \#coronavirus, \#BlackLivesMatter, and \#Christmas}.

Interestingly, focus Inflammatory and Insult profiles share hashtags about American political situation in a cohesive set of hashtags such as \emph{\#Trump, \#DumpTrump, \#TrumpIsALoser}, as well as \emph{\#MAGA} which points to the campaign led by Trump and its supporters in the last 6 years.
Hashtags like \#FakeNews, \#MeTooMovement and \#NBA are used in aggravating manner as well.

\textbf{Takeaways 3-4:}
First, a minority of focus profiles use hashtags; when used, they are sparse, compared to the volume of unique hashtags from equal number of random profiles.
Second, hashtags shared by focus profiles are coherent, but toxic and malignant in nature, compared to random.

\begin{table}[t]
\centering
\small
\tabcolsep=0.05cm
\scalebox{0.78} {
\begin{tabular}{l | r r | r r | r r | r r  }
\toprule

         & \multicolumn{2}{c}{\bf Identity Attack} & \multicolumn{2}{c}{\bf Inflammatory}  & \multicolumn{2}{c}{\bf Insult} & \multicolumn{2}{c}{\bf Threat}\\
             \cline{2-9}

& \mc{1}{Focus} &{Random}   & \mc{1}{Focus} & {Random} & \mc{1}{Focus} & {Random} & \mc{1}{Focus} & \mc{1}{Random}\\
\midrule
\# Users		&	65	&	65	&	161	& 161	&	52	&	52	&	34	&	34	\\
\# All HTags		&	701	&	9,337	&	5,299	&	35,071	&	771	&	7,728	&  1,426	&	5,579\\
\# Unique HTags		&	612	&	7,741	&	4,008	&	24,627	&	742	&	6,545	&	1,355	&	4,787	\\
Avg. All HTags	&	11	&	145.9	&	33.5	&	222	&	16.1	&	161		&	44.6	&	174.3\\
Avg. Uniq HTags	&	9.6	&	121	&	25.4	&	155.9	&	15.5	&	136.4		&	42.3	&	149.6\\
\bottomrule
\end{tabular}
}
\caption{Breakdown of hashtags for focus vs. random profiles, across four categories of misbehavior.
}
\label{tab:thags}
 \vspace{-0.3cm}
\end{table}

\begin{table}[t]
\centering
\small
\tabcolsep=0.05cm
\resizebox{1.0\columnwidth}{!}{
\begin{tabular}{r | l }
\toprule
{\bf Category} & {\bf Top 5 Hashtags} \\ \midrule
Identity Attack (Focus)     &   TreCru (46.9\%), Syria (3.1\%), BDS (3.1\%),Germany (3.1\%), Jesus (3.1\%)\\ \cline{2-2}
\multirow{2}{*}{(Random)}   &   COVID19(37.5\%), coronavirus (31.3\%), BREAKING (26.6\%), \\
& BlackLivesMatter (31.325\%), Christmas (21.9\%)\\\midrule
Inflammatory (Focus)        &   MeToo (10.8\%), metoo (8.9\%), Trump (7.6\%),	MAGA (7.6\%), FakeNews (7\%) \\ \cline{2-2} 
\multirow{2}{*}{(Random)}                     &   COVID19 (36.1\%), coronavirus (22.8\%),	BlackLivesMatter (21.5\%),
\\ 
& BREAKING (21.5\%), Trump (14.6\%) \\\midrule
\multirow{2}{*}{Insult (Focus)}              &   trap (4.2\%), NBA (4.2\%), DumpTrump2020 (4.2\%), \\
& TrumpIsALoser (4.2\%), cum (4.2\%) \\ \cline{2-2} 
\multirow{2}{*}{(Random)}                    &   COVID19 (45.8\%), BlackLivesMatter (43.8\%),	coronavirus (31.3\%), \\
& BREAKING (29.2\%), Trump (25\%) \\\midrule
Threat (Focus)              &   urgent (6.3\%),	pixelgun (6.3\%), fps (6.3\%), pg3d (6.3\%), mobile (6.3\%) \\ \cline{2-2} 
\multirow{2}{*}{(Random)}                    &   COVID19 (34.4\%), BREAKING (31.3\%), TBT (21.9\%), \\
& coronavirus (21.9\%), GRAMMYs (18.8\%) \\ \bottomrule
\end{tabular}
}
\caption{Top 5 hashtags in four categories of misbehavior, compared to random profiles. Note that (\%) shows the percentage of selected users using a given hashtag.}
\label{tab:top5_htags_gini_median}
\vspace{-0.2cm}
\end{table}

\subsection{How diverse are the topics of tweets?}
\label{sec:topic-diversity}
\begin{table}[t]
\centering
\tabcolsep=0.05cm
\scalebox{0.60} {
\begin{tabular}{l | c | l  }
\toprule

{\bf Category} & {\bf Topic} & {\bf Top 5 words} \\
\toprule
\multirow{3}{*}{\vtop{\hbox{\strut Identity attack}\hbox{\strut (Focus)}}}	&	1	&	occupier(0.15), israeli(0.12), occupy (0.08),	palestinian (0.08), attack (0.07)	\\
	&	2	&	fuck(0.45), tit(0.08), petit(0.08), wear(0.07),	chechick(0.06)	\\
	&	3	&	suck(0.15), anal(0.13),	fucking(0.11), busty(0.10), enjoy(0.09)	\\ \midrule
\multirow{3}{*}{\vtop{\hbox{\strut Identity Attack}\hbox{\strut (random)}}}	&	1	&	know(0.27), happy(0.12), big(0.11), morning(0.10), hear(0.07)	\\
	&	2	&	love(0.24), people(0.19), use(0.12), free(0.07), read(0.06)	\\
	&	3	&   amp(0.23), time(0.14), feel(0.09), talk(0.09), hope(0.08)	\\\midrule
\multirow{3}{*}{\vtop{\hbox{\strut Inflammatory}\hbox{\strut (Focus)}}}	&	1	&  people(0.26), make(0.20), tell(0.12), trump(0.09), stop(0.08)	\\
    &	2	& need(0.23), vote(0.13),free(0.11), israeli(0.92), home(0.76)	\\
	&	3	& child (0.16), woman (0.16), person (0.16), antebellum (0.09), human (0.07)	\\\midrule
\multirow{3}{*}{\vtop{\hbox{\strut Inflammatory}\hbox{\strut (Random)}}}	&	1	& day (0.25),year(0.18),good(0.15),amp(0.07),man(0.07)	\\
    & 2 & say (0.27), people (0.16), vote (0.08), team (0.08), check (0.07)	\\
    & 3 & know(0.23), need (0.17), thing (0.12), let (0.10), tell (0.08)	\\\midrule
\multirow{3}{*}{\vtop{\hbox{\strut Insult}\hbox{\strut (Focus)}}}	&	1	&	ass (0.20), girl (0.17), fuck (0.15), want (0.09), say (0.08)	\\
	&	2	&	shave (0.17), blonde (0.13),	pussy (0.11),	wear (0.10), babe (0.09)	\\
	&	3	&	teen (0.19), sex (0.16),	rebel (0.09),	misfit (0.08),	round (0.08)	\\\midrule
\multirow{3}{*}{\vtop{\hbox{\strut Insult}\hbox{\strut (Random)}}}	&	1	&	people (0.21), think (0.19),	really (0.12),	history (0.07), today (0.07)	\\
	&	2	&	say (0.20),	need (0.15), watch (0.09), let(0.09), black (0.09)	\\
	&	3	&	amp (0.28),	driver (0.13),	make (0.09), day (0.08), play (0.08)	\\\midrule
\multirow{3}{*}{\vtop{\hbox{\strut Threat}\hbox{\strut (Focus)}}}	&	1	& 	black (0.12), park (0.11), kill (0.09), weed (0.07), gay (0.06)	\\
	&	2	& kill (0.22), playhit (0.18), shell(0.16), warplane (0.13), man (0.08)	\\
	&	3   &  high (0.70), assault (0.14), common (0.14), rise(0.01), park (0.01) \\ \midrule
\multirow{3}{*}{\vtop{\hbox{\strut Threat}\hbox{\strut (Random)}}}	&	1	& video (0.21), good (0.17), day (0.17), like (0.14), look (0.09)	\\
	&	2	& amp (0.24), make (0.21), check (0.11), post (0.07), hope (0.68)	\\
	&	3	& new (0.30), love (0.24), vote (0.10), shit (0.10), night (0.08)	\\
\bottomrule
\end{tabular}
}
\caption{Top 3 topics from LDA analysis of tweets of focus and random profiles, in four categories of misbehavior (due to space, only top 5 words shown per topic).}
\label{tab:top5_htags_gini_sum_2}
\vspace{-0.2cm}
\end{table}

The text in a tweet is of great value to peek into the nature of one-way posting and even discussions a profile engages in.
In order to generalize the cohesiveness and types of topics discussed in our focus profiles (which are representative of the four categories of misbehavior), we use Topic modeling on the text of all tweets in each set of profiles.

We specifically used Latent Dirichlet Allocation (LDA)~\cite{10.5555/944919.944937}, a probabilistic topic modeling algorithm to extract topics from a number of documents (user tweets).
The LDA assumes that each document is composed of a number of topics, with each topic existing as a probability distribution over all the words in the topic.
Our implementation to extract topics from users' tweets in toxic and random profiles leverages the Natural Language Toolkit~\cite{NLTK}.
For each profile in focus or random sets, we extract its tweets' text.
We remove the re-tweets' text, and consider English-only tweets.
We then remove all non alpha-numeric characters, URLs, and stop-words, and then apply lemmatization and tokenization.
Finally, we extract topics for each category of misbehavior and random profiles.

Table~\ref{tab:top5_htags_gini_sum_2} shows the top 3 topics (via the top 5 highest weighted words for each) found in the four categories of focus profiles, as well as random.
We observe that the topics extracted from the sets of focus profiles are consistent and less diverse.
Topics in focus Identity attack profiles are about Israel and Palestinian war, with words like occupier, israeli, Zionist, and attack, suggesting blunt and forceful language used in those tweets.
Focus Inflammatory profiles talk about politics, using words such as trump, vote, people, antebellum, etc.
The focus Insult profiles talk about explicit, pornographic-related topics.
As discussed earlier in $\S$~\ref{sec:remove_obscene}, pornographic content is still captured by high Perspective model scores, with topic \#2 and \#3 containing strong cursing, swearing and profane language.
Focus Threat profiles exclusively talk about fight and war and other violent topics such as assault, kill, and playhit.
Interestingly, all these topics are in stark contrast to the topics of random profile sets that are found, as perhaps expected, to have no cohesiveness and covering a very diverse spectrum such as love, happiness and feelings, history, voting, teams, people, etc.

We also extract topics from the retweets of focus and random profiles, and discover similar results.
For example, focus Inflammatory profiles retweet about Trump and women, and focus Insult profiles retweet about Black Lives Matter movement and killing.
Contrasting these topics, all random groups shared in-cohesive benign topics.

\textbf{Takeaways 5-6:}
First, topics of discussion of each cluster are related to the type of misbehavior the cluster represents.
Second, there is cohesiveness in the theme of topics within the focus profiles of a cluster.
Focus profiles, in general, discuss topics that project hatred, insult, threat and sensitive topics about war zones and politics, while random profiles discuss harmless topics, such as users' daily lives and feelings, history or books.
As a whole, focus Identity Attack, Inflammatory, Insult and Threat profiles consistently share very specific, hateful, sensitive and obscene-natured content.
\subsection{How readable are the tweets posted?} 
\label{sec:readability}

We further analyze our focus and random profiles' tweets for grammatical and semantic correctness.
We parse each tweet to extract the number of words, sentences, punctuation, non-letters (e.g., emoticons), and measure the \emph{Lexical Richness}, the \emph{Automated Readability Index (ARI)}~\cite{senter1967automated} and the \emph{Flesch Score}~\cite{flesch1948new}.
Lexical richness, defined as the ratio of number of unique words to total number of words, reveals noticeable repetitions of distinct words.
ARI estimates the comprehensibility of a text corpus and is computed as: (4.71$\times$average word length)+(0.5$\times$average sentence length)-21.43.
Flesch score indicates how difficult it is to read the text and is computed as: 206.835-1.015$\times$($\frac{total words}{total sentences}$)-84.6$\times$($\frac{total syllables}{total words}$).
Higher value of ARI, and higher Flesch score of a given text show its comprehensiveness and easy readability. 
Table~\ref{tab:lanalysis} shows a summary of the results.
In comparison to focus profiles, random profiles tweet with higher ARI compared to Identity Attack focus profiles (9.08 vs. 7.73), higher Richness (0.22 vs. 0.10), and higher Flesch score (58.66 vs. 53.42) 
thus suggesting that non-toxic profiles use a richer vocabulary, and that their tweets have higher readability.

\textbf{Takeaway 7:}
Focus profiles tweet with lower comprehensibility and readability, and poorer vocabulary than random.

\begin{table}[t]
\centering
\small
\tabcolsep=0.05cm
\resizebox{1.0\columnwidth}{!}{
\begin{tabular}{l | r r | r r  | r r  | r r  }
\toprule

    & \multicolumn{2}{c}{\bf Avg. \# of Sentences}  & \multicolumn{2}{c}{\bf Avg. Richness}  & \multicolumn{2}{c}{\bf Avg. Flesch Score} & \multicolumn{2}{c}{\bf Avg. ARI} \\
            \hline
Categories      &	Focus	&	Random &	Focus	&	Random	&	Focus	&	Random	&	Focus	&	Random	\\
	\hline
Identity Attack	&		2885.46	&	1416.52 & 0.10	&	0.22
	&	53.42	&	58.66	&	7.73	&	9.08	\\
Inflammatory	&		1849.28	&	1350.51	& 0.19	&	0.22
& 	56.07	&	60.31	&	6.75	&	7.71	\\
Insult	&		2616.08	&	1519.25 & 0.14	&	0.20
	&	52.90	&	59.68	&	6.37	&	8.88	\\
Threat	&		1515.74	&	1301.32 & 0.16	&	0.22
	&	42.07	&	64.55	&	5.69	&	6.81	\\

\bottomrule
\end{tabular}
}
\caption{Overview of lexical analysis on tweet content. 
}
\vspace{-0.2cm}
\label{tab:lanalysis}
\end{table}

\section{Content Homogeneity of Focus Profiles}
\label{sec:homogeneity}

We have seen that focus profiles share less diverse content in terms of URLs, web domain categories, hashtags, and text-based topics.
To discern if tweets of these profiles are homogeneous between profiles (and potentially working for the same influence operation) of a given selection, we further investigate the Jaccard similarity between domains (\S\ref{sec:similarity-domains}) and hashtags (\S\ref{sec:similarity-hashtags}) in tweets, within and between their respective clusters.
We also study these similarities on all four types of misbehavior and compare how their distributions differ by computing their KL-Divergence distance.


\subsection{How homogeneous are the domains shared?}
\label{sec:similarity-domains}

We assess the overlap among the domains referenced in tweets of focus and random profiles using the pairwise Jaccard similarity index, computed between domains lists $A$ and $B$ referenced by focus Identity Attack or random profiles, respectively.
The \emph{Jaccard Index} is computed for two sets $A$ and $B$ as $\frac{|A \bigcup B|}{|A \bigcap B|}$, and ranges from 0 (for no common elements between the two sets) to 1 (for a perfect match or overlap).
Similar results are obtained for the other dimensions and are omitted for brevity.
Figures~\ref{fig:url_domain_puser:gini_median-dom-simm} and ~\ref{fig:url_htags_similarity:gini_median-simm} present the CDF of the Jaccard similarity scores of focus and random profiles, as obtained for domains and hashtags in their tweets, respectively.
As expected, the great majority of random profiles are highly dissimilar with each other, as well as with focus profiles (Figure~\ref{fig:url_domain_puser:gini_median-dom-simm}: $\sim$1.3\% of pairs of random profiles have max 0.2 similarity).
However, within focus profiles, $\sim$50\% of profiles have no similarity with each other, $\sim$25\% have up to 0.3 similarity, and the rest have similarity higher than 0.3.
Also, only 1-2\% of focus profiles show any similarity with random profiles, indicating that the messages or objective of the focus profiles largely differ from random.

\begin{figure*}[t]
    \centering
   
    \subfloat[{\small Number of domains}]{
    \includegraphics[width=0.50\columnwidth]{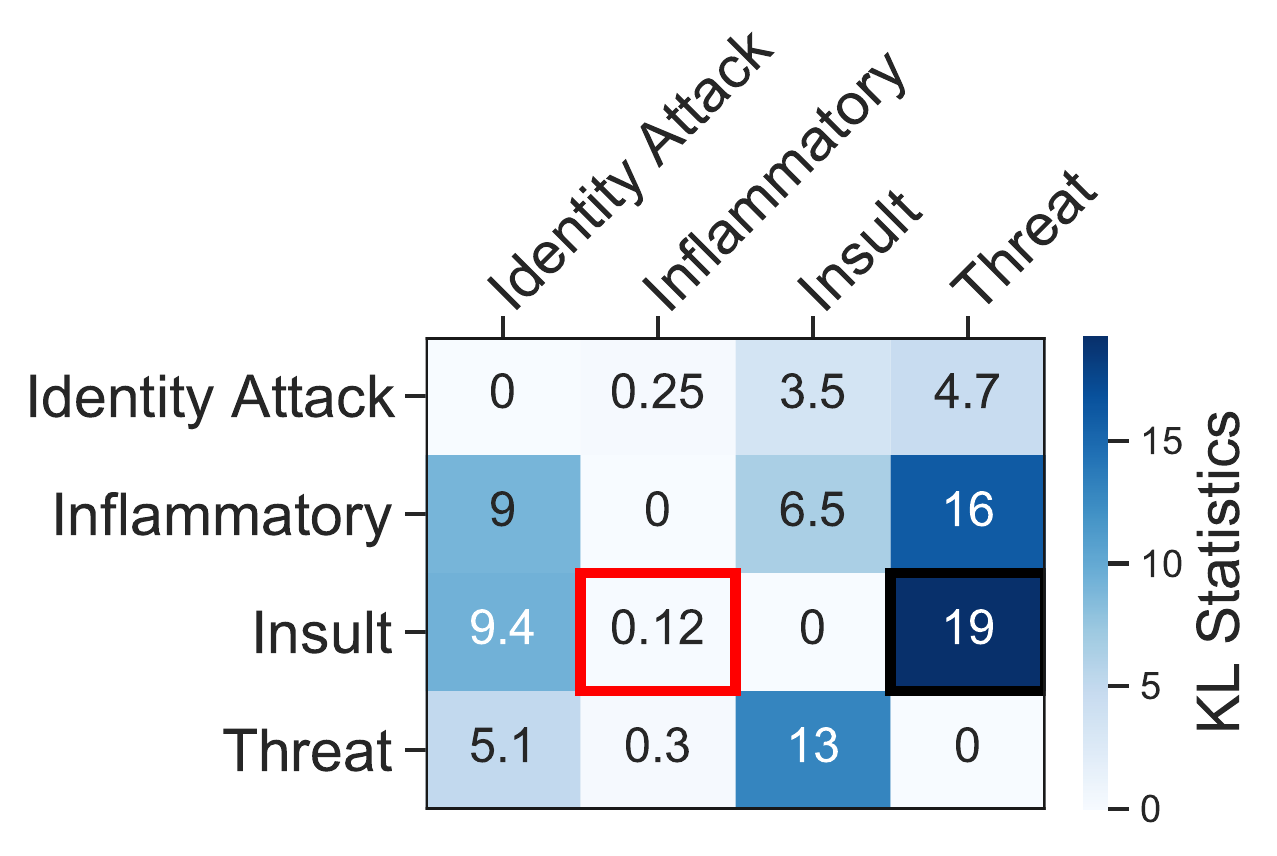}\label{fig:kl_domain_url}
    }
    \subfloat[{\small Similarity of domains}]{
    \includegraphics[width=0.50\columnwidth]{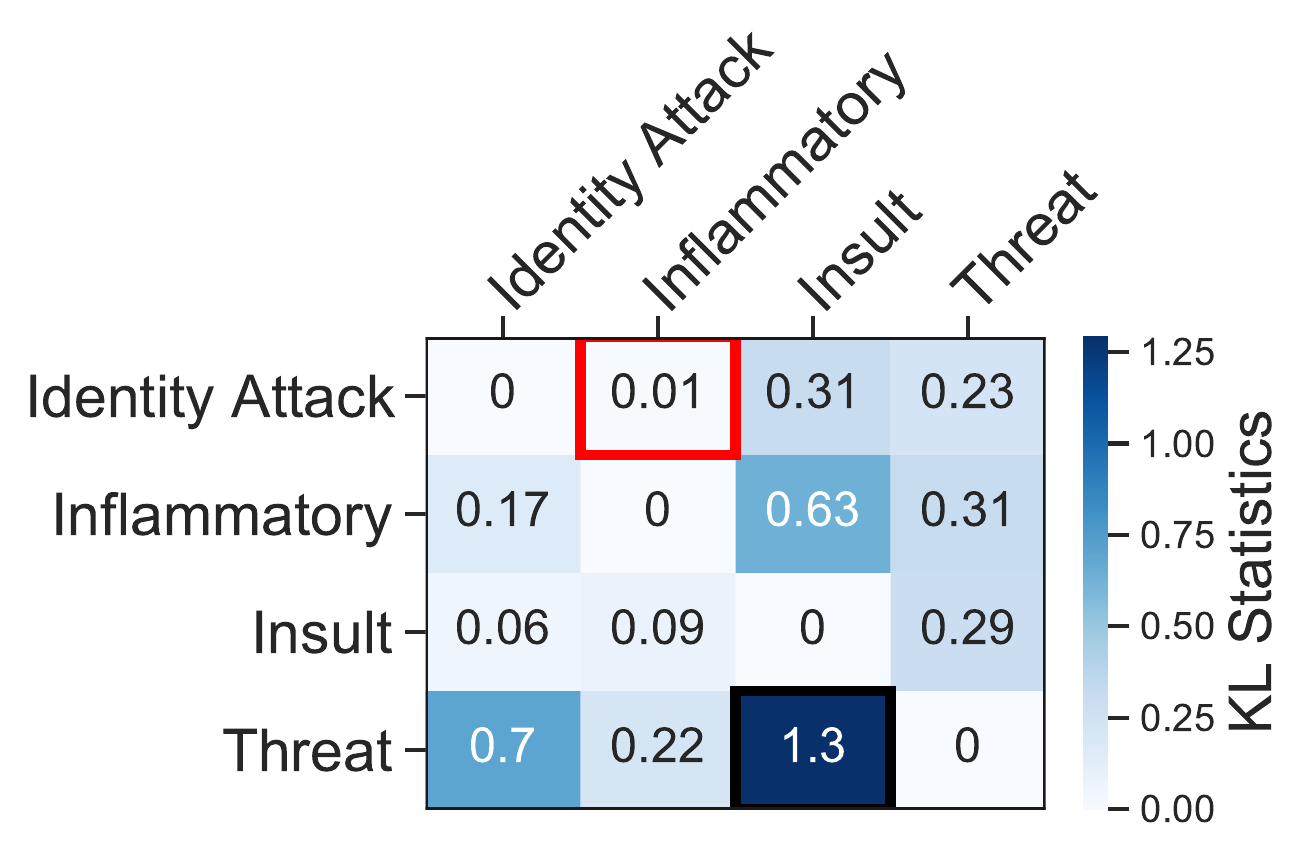}\label{fig:kl_domain_sim}
    }
    \subfloat[{\small Number of hashtags}]{
    \includegraphics[width=0.50\columnwidth]{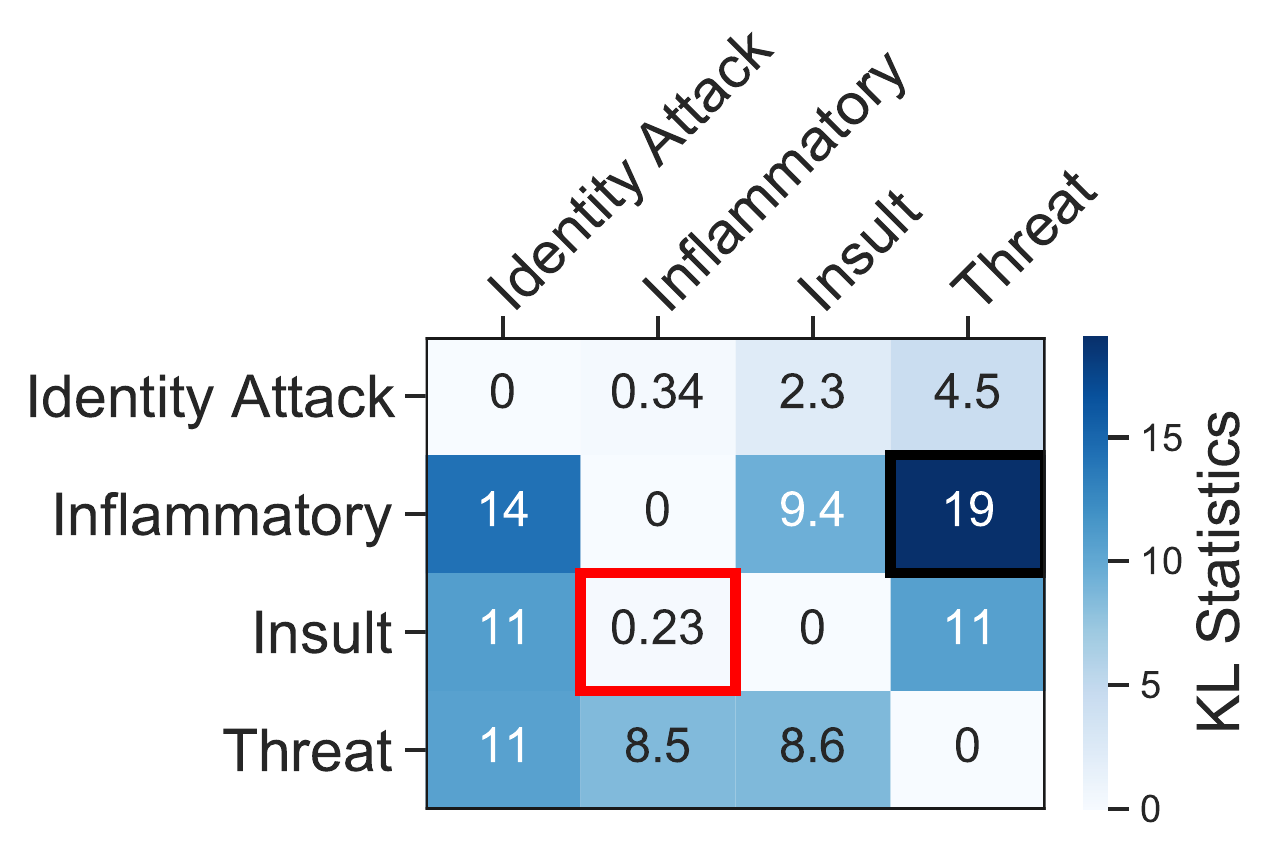}\label{fig:kl_hashtag_dist}
    }
    \subfloat[{\small Similarity of hashtags}]{
    \includegraphics[width=0.50\columnwidth]{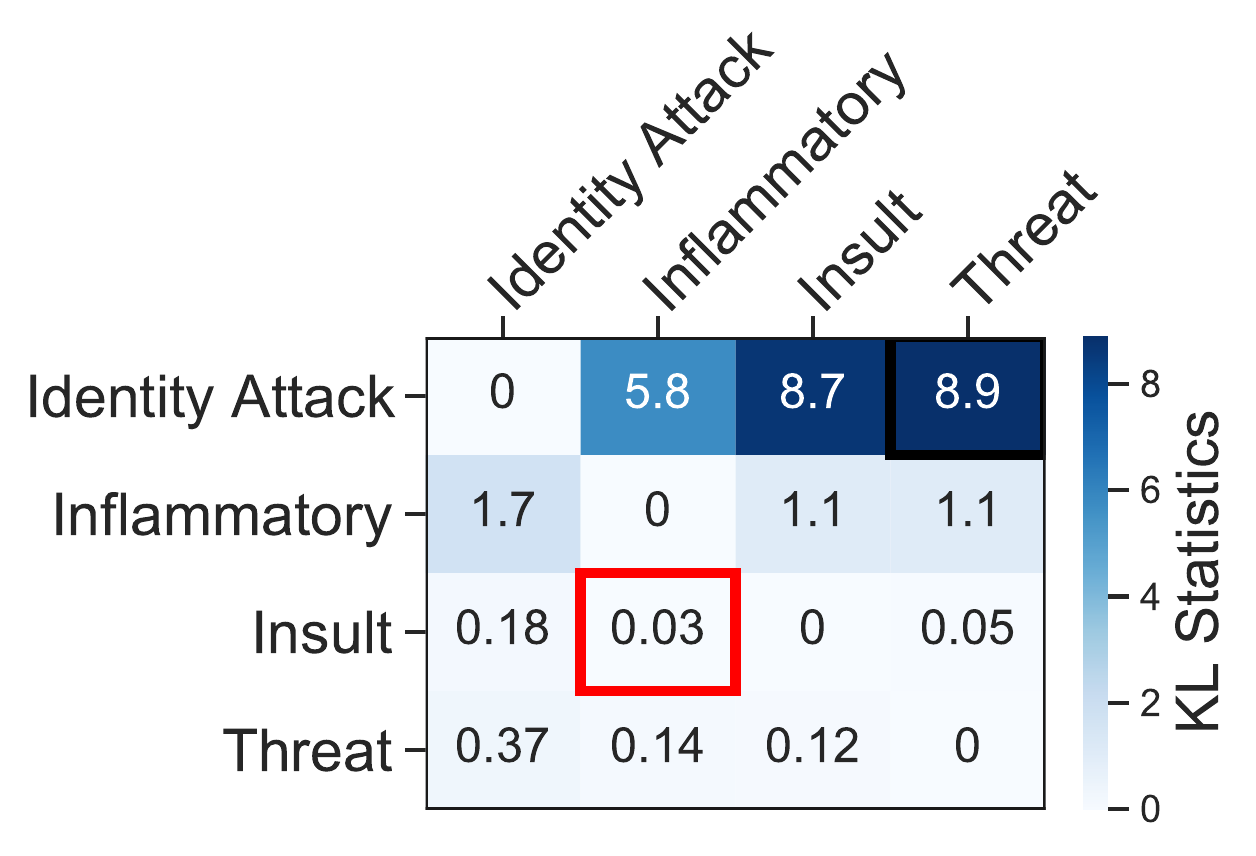}\label{fig:hashtag_sim}
    }
    \caption{Heat maps of Kullback-Leibler (KL) divergence ($D_{KL}$) between consistently toxic profiles.
    $D_{KL}$ is calculated for CDFs of (a) number of domains, (b) similarity of domains, (c) number of hashtags, and (d) similarity of hashtags for four types of focus profiles.
    Lower $D_{KL}$ (indicated by white color) points to more similar toxic profiles for that pair measurement.
    Red squares: Insult \& Inflammatory focus profiles have very similar distributions of number of domains, hashtags and hashtag similarity; Identity Attack \& Inflammatory focus profiles have high domain similarity.}
\label{fig:kl_plots}
\vspace{-0.2cm}
\end{figure*}

Table~\ref{tab:domain_url_cat_stats} showed the number of domains and their categories.
To assess the distance between their distributions, we compute the \emph{Kullback–Leibler Divergence (KL)} between the number of domains in all four sets of focus profiles.
$D_{KL}(P||Q)$ is a statistical distance of how one probability distribution function $P$ is different from a second one $Q$.
In Figure~\ref{fig:kl_domain_url}, we observe that the number of domains shared in focus Insult profiles is closest to focus Inflammatory ($D_{KL}$:0.12), whereas the number of domains in focus Identity Attack and Threat profiles are also closest to Inflammatory profiles ($D_{KL}$:0.25, 0.3). 
We also measure $D_{KL}$ between all pairs of random profile sets and find they differ greatly ($D_{KL}$:36.27-49.39, details omitted due to space).

To analyze the homogeneity of domains found in tweets of profiles belonging to different categories of misbehavior, we also compute $D_{KL}$ for the similarity of the domains (cf. Figure~\ref{fig:url_domain_puser:gini_median-dom-simm}). 
Figure~\ref{fig:kl_domain_sim} shows that the similarity of domains referred in URLs of focus Identity Attack, Insult and Threat profiles have minimum $D_{KL}$ from focus Inflammatory profiles ($D_{KL}$: 0.01, 0.09, 0.22), pointing to the fact that domains are very similar.
Also, focus Identity Attack domains are very similar to the ones shared in focus Insult and Threat ($D_{KL}$: 0.31, 0.23).
In contrast, random profiles show higher $D_{KL}$ scores for similarity among domains, suggesting the use of distinct domains in tweets of random profiles.

\textbf{Takeaway 8:}
Focus profiles across all toxic categories, in comparison to random, are homogeneous in terms of the number of, and actual domains they post in their tweets.

\subsection{How homogeneous are the hashtags shared?} 
\label{sec:similarity-hashtags}

Following the analysis on domains, we computed Jaccard similarity on the sets
of hashtags, as well as $D_{KL}$, as measures to gauge similarity among sets of focus profiles.
Figure~\ref{fig:url_htags_similarity:gini_median-simm} shows the CDF of the Jaccard similarity computed between the vectors of hashtags appearing in tweets of focus Identity Attack profiles, between random profiles, and across focus and random profiles.
Similar results were retrieved for the other dimensions and are excluded due to space.
We find that focus profiles within the specific cluster are more similar with respect to usage of hashtags, than random profiles.
In particular, $\sim$86\% of the focus Identity Attack profiles have at least 0.5 Jaccard similarity, and
compared to the similarity within random profiles, where only 5\% of pairs have a similarity $\geq$0.5.
Also, focus profiles are using distinctly different hashtags from random, since their cross-profile similarity is close to 0 in 97\% of cases.

We now turn our attention to the number of hashtags used by each profile.
To calculate $D_{KL}$ of distribution of hashtags used per cluster of focus profiles, we first computed the CDFs of total hashtags per profile in all four sets of focus profiles. 
Figure~\ref{fig:kl_hashtag_dist} shows that the lowest $D_{KL}$ values are found for Inflammatory vs. Identity Attack profiles (0.34), and Inflammatory vs. Insult (0.23) profiles, indicating that they share similar number of hashtags.
High $D_{KL}$ scores for random sets of profiles showed that these profiles, as earlier found, have quite different distributions with each other (min $D_{KL}$ in random set comparisons is 12.4).
In Figure~\ref{fig:hashtag_sim}, the lowest $D_{KL}$ values show that Insult, Threat, and Inflammatory profiles share most similar hashtags.
Again, high $D_{KL}$ of hashtags similarity among random profiles showed the nature of hashtags they share does not match to each other.

\textbf{Takeaway 9:}
Focus profiles across all toxic categories, in comparison to random, are homogeneous in terms of the number of, and actual hashtags they post in their tweets.

\section{Posting Automation in Focus Profiles}
\label{sec:time-analysis}

Twitter provides quick updates about significant events happening around the world.
Unfortunately, many of these updates are in part due to automatic accounts with 66\% of tweeted links from popular news and current event websites are made by Twitter Bots~\cite{PewResearchCenter2018}.
Beyond news, bots have been scrutinized for also spreading fake news and changing public political perception and discourse in coordinated influence operations.
Our results on extracted topics, shared hashtags and URLs on the identified focus profiles provide hints to the possible automation of these accounts, in line with~\citet{PewResearchCenter2018}.
Thus, we now attempt to characterize how many of the focus profiles, which are consistently posting very specific and toxic content, could be bots.
Following past work by~\citet{9108896}, we analyze their tweeting time patterns to infer periodic and bot-like behaviors, and compare it with random profiles (\S\ref{sec:tweeting-patterns}).
We also query \emph{Botometer} by~\citet{botometer} for a likelihood that a given profile is bot (\S\ref{sec:botometer}).

\begin{figure*}[t]
    \begin{subfigure}[t]{0.22\linewidth}
            \includegraphics[width=\textwidth]{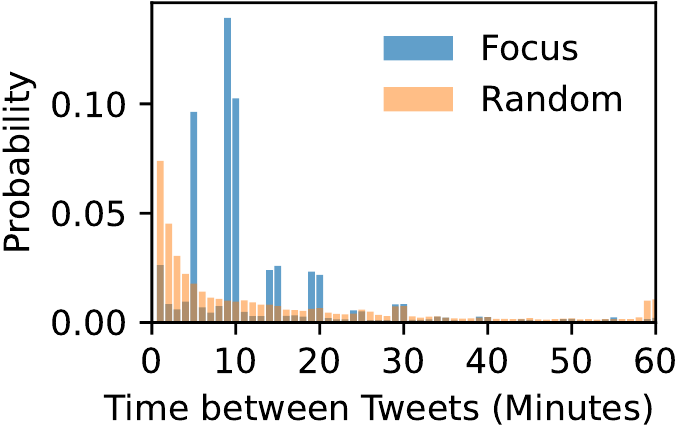}
            \caption{Inter-tweet intervals}  \label{fig:tbtm_gini_median}
    \end{subfigure}\hfill
      \begin{subfigure}[t]{0.22\linewidth}
            \includegraphics[width=\textwidth]{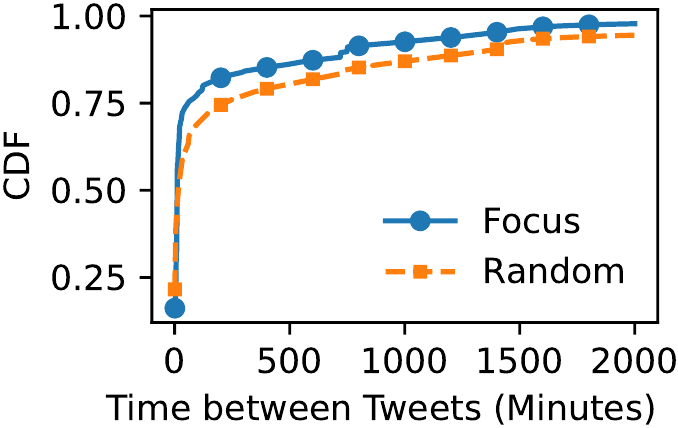}
            \caption{Inter-tweet intervals}
    \label{fig:tbtcdf_gini_median}
    \end{subfigure}
    \hfill
    \begin{subfigure}[t]{0.22\linewidth}
        \includegraphics[width=\textwidth]{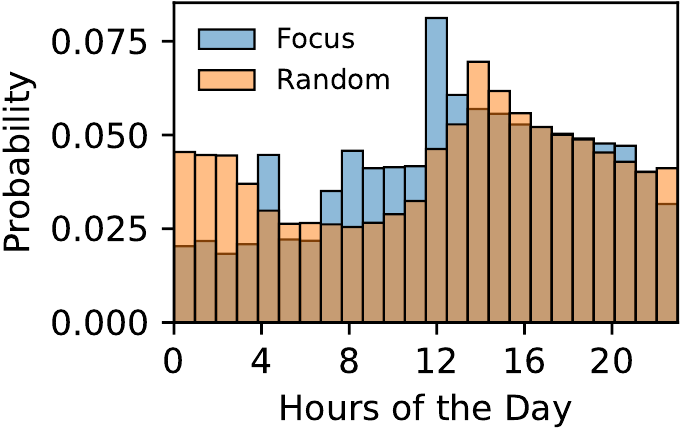}
            \caption{Tweets per hour}
            \label{fig:tbth_gini_median}
    \end{subfigure}
    \hfill
    \begin{subfigure}[t]{0.22\linewidth}
            \includegraphics[width=\textwidth]{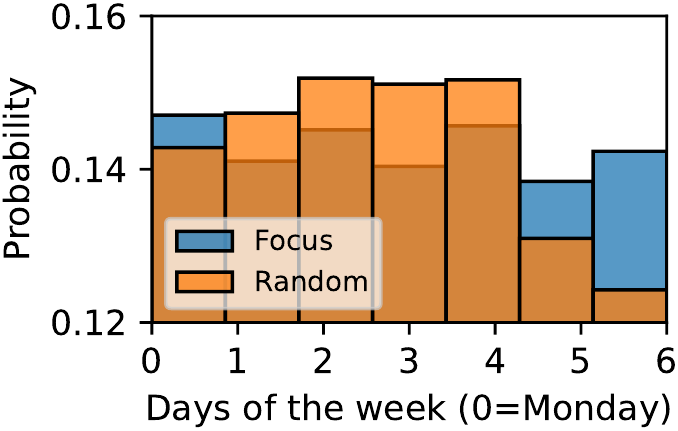}
            \caption{Tweets per day}
            \label{fig:tbtw_gini_median}
    \end{subfigure}
    \caption{
    Overview of temporal analysis: (a) PDF of tweet inter-arrival time;
    (b) CDF of tweet inter-arrival time;
    (c) PDF of tweet posting time during the day;
    (d) PDF of tweet posting time during the week.
    }
    \label{fig:day_week_gini_median}
    \vspace{-2mm}
\end{figure*}

\subsection{How regular is the tweeting pattern in time?}
\label{sec:tweeting-patterns}

We define \emph{tweeting pattern} as the frequency and timing of a profile's tweets.
In order to investigate the tweeting pattern of focus profiles, we first isolate the timestamps of all their tweets.
Figure~\ref{fig:tbtm_gini_median} shows the Probability Distribution Function (PDF) of time between sequential tweets by focus Identity Attack and random profiles, up to 60 minutes (similar results were retrieved for the other clusters and can be found in Appendix Sec. ~\ref{A:coordination analysis}).

We observe that these focus profiles produce tweets at highly regular intervals, with clear peaks at $<$1, 5, 10, 15, and 20 minutes, with 75.2\% of all inter-tweet intervals occurring faster than an hour.
On the other hand, random profiles have a smooth distribution of inter-tweet intervals, producing new tweets in all possible time slots, and almost monotonically decreasing as the interval increases.
Also, 64.3\% of all inter-tweet intervals occur within an hour.
Complementing Figure~\ref{fig:tbtm_gini_median}, Figure~\ref{fig:tbtcdf_gini_median} shows the CDF of inter-tweet intervals: $20\%$ of focus~(random) profile tweets have an inter-arrival time larger than 123~(463) 
minutes, with a maximum of inter-arrival time of 22~(12) days. 

Then, we look into the time of day and day of week that tweets are being posted by focus or random profiles.
Figures~\ref{fig:tbth_gini_median} and~\ref{fig:tbtw_gini_median}  show the PDFs for these two tweeting pattern aspects.
We observe that the focus profiles are more active starting from UTC 7am to 2pm than random profiles, which observe the typical diurnal behavior of regular user, with dual peaks during Americas and European working and evening hours.

Further, focus profiles are quite different in their posting activity during the week than random profiles: they maintain similar levels of activity throughout the whole week, and are more and consistently active on weekends, compared to random profiles who demonstrate a notable decrease in their weekend activity.

When we repeat this analysis on the remaining types of misbehavior (for each set of toxic and random profiles) we observe similar results, rest of the plots can be found in Appendix (Sec.~\ref{A:coordination analysis}).
We also elaborate the results as a $D_{KL}$ score to the Identity Attack analysis.
The $D_{KL}$ of the inter-tweet time distribution for Identity Attack (i.e., Figure~\ref{fig:tbtm_gini_median}) to Inflammatory, Insult and Threat focus profiles is $D_{KL}$:~0.40,~0.01,~0.04, respectively.
Similarly low scores are found when computing $D_{KL}$ on the CDFs (i.e., Figure~\ref{fig:tbtcdf_gini_median} for Identity Attack), etc.: $D_{KL}$:~0.01,~0.01,~0.04, respectively.

\textbf{Takeaway 10:}
The tweeting behavior of focus profiles is very consistent and regular.
They tweet frequently and at specific, small time intervals, and demonstrate longer activity hours during the day and week, without the typical breaks during weekend, that random profiles show.
\subsection{How bot-like are toxic profiles?}
\label{sec:botometer}

To complement our findings, we also query \emph{Botometer} by~\citet{botometer} for a score for all focus and random profiles.
Botometer is an AI-based algorithm that classifies a given Twitter account as bot/automated account or human.
It takes into account a profile's followers, friends, account age, sentiment and language of its tweets, and outputs a \emph{bot score} ranging from 0 to 5, with 0 being most human-like and 5 being the most bot like.
The received Botometer scores are shown in Table~\ref{tab:botometer}, for all four misbehavior dimensions and for focus vs. random profiles.
We find that our focus profiles score higher than random profiles in being bots, in a consistent manner, and their scores are equally tight with random profiles in score variability.
We also retrieve the Complete Automation Probability (CAP) by~\citet{yang2019arming}, a conditional probability of a profile being a bot with a given Botometer score.
For example, Focus Identity Attack profiles have an average 4.53 Botometer score; at this score, 89.7\% of accounts with this score or higher are likely to be bots, in contrast to the 57.8\% of accounts for random profiles with 1.58 Botometer score.

\textbf{Takeaway 11:}
Focus profiles demonstrate characteristics that rank them higher in the bot scale of Botometer, hinting to the higher likelihood they are automated accounts.

\begin{table}[t]
\centering
\resizebox{1.0\columnwidth}{!}{
\begin{tabular}{l|cc|cc}
\toprule
       & \multicolumn{2}{c|}{Focus}             & \multicolumn{2}{c}{Random}            \\ \cline{2-5}
       & Bot Score         & CAP (\%)               & Bot Score         & CAP (\%)              \\ \midrule
Identity Attack & 4.531 $\pm$ 0.64 & 89.7 $\pm$ 5.8 & 1.579 $\pm$ 1.60 & 57.8 $\pm$ 25.4 \\
Inflammatory    & 3.281 $\pm$ 1.38 & 81.3 $\pm$ 9.1 & 1.358 $\pm$ 1.52 & 54.6 $\pm$ 26.1 \\
Insult          & 4.038 $\pm$ 1.08 & 85.3 $\pm$ 9.4 & 1.645 $\pm$ 1.60 & 60.1 $\pm$ 24.7 \\
Threat          & 4.096 $\pm$ 1.13 & 84.9 $\pm$ 9.6 & 1.278 $\pm$ 1.37 & 56.1 $\pm$ 24.4 \\ \bottomrule

\end{tabular}
}
\caption{Mean and Standard Deviation of Botometer scores and Complete Automation Probability (CAP) for selected focus profiles vs. random profiles.
}
\label{tab:botometer}
\vspace{-2mm}
\end{table}

\section{Related Work}
\label{sec:rwork}

Online misbehavior detection on social networks has been extensively explored by several studies such as~\citet{gomez2019exploring,ribeiro2018like,founta2018large, waseem-hovy-2016-hateful, DHUNGANASAINJU2021106735}, to name a few.
This work identifies different types of bullying and online misbehavior and derives user motivations behind users involved in bullying on Twitter, along with an examination of temporal patterns in bullying-related tweets.
The past studies have availed human annotations to differentiate between toxic and non-toxic tweets.
This work relies on ML models of Perspective API to rate the collected tweets.
This work also explored misbehavior dimensions beyond the prior works, and at larger scale of data, i.e., 293M tweets.

\citet{hosseini2017deceiving} and \citet{jain2018adversarial} have studied the Google's Perspective API~\cite{perspective} and its resilience against adversarial attacks.
Those studies leveraged Perspective API to score and analyze the toxicity of tweets.
This work takes precedence over these studies in terms of size of the data set (293M tweets) and the number of misbehavior dimensions not studied in the past, namely Insult, Inflammatory, Threat, and Identity Attacks.

Influence operations on OSNs is a heightened phenomenon, which spreads through automated accounts or bots~\cite{PewResearchCenter2018}.
Consistent toxic and false content creation and dissemination is the base of a consistent spread of toxicity on OSNs by active operators or accounts~\cite{FFI-RAPPORT}. Content-based features best predict OSN-based influence operations, but unsupervised ML for detection of coordinated efforts of profiles in carrying these operations are infeasible at scale~\cite{doi:10.1126/sciadv.abb5824}.
This longitudinal study of 14 years gives a very clear picture of consistent production of toxic content.
The presented methodology effectively differentiates consistent vs. occasional misbehavior of Twitter profiles, and allows spotting the consistently malignant content and profiles.

\section{Discussion and Conclusion}
\label{sec:conclusion}

\textbf{Summary:}
In this paper, we performed a first of its kind longitudinal study of 122K Twitter profiles and 293M tweets, over a period of 15 years (2007-2021).
We were particularly interested in studying toxic profiles who may participate in influence operations on Twitter.
Towards this goal, we analyzed the toxicity of tweets using six Perspective API ML models and found that toxic behavior has increased through this 15-year period, across all six dimensions of misbehavior.
We took a deep dive into the most toxic profiles, who are also very consistent in this behavior.
We focused on these profiles and studied their posted content, topics covered and post timing patterns, and observed several characteristics that can help identification and removal from OSNs.
These focus profiles are noticeably different to random Twitter profiles in terms of shared content and posting patterns.

\textbf{Findings on consistent, and highly toxic profiles:} 
\vspace{-2mm}
\squishlist
\item They fetch and share very specific and cohesive in type web resources (domains), originating from many URLs.
\item They post from a homogeneous and small in size pool of domains shared within their cluster of misbehavior.
\item Less than 1/3rd of them use hashtags; and their hashtags are mostly malignant and toxic in nature. 
\item They tweet on topics that are cohesive and related to their type of misbehavior: hatred, insult, threat, and sensitive topics about war zones and politics.
\item Their text has lower comprehensibility and readability, and uses poorer vocabulary than random profiles.
\item They tweet in small and regular time intervals, and often coincide with each other's posting activity.
\item They demonstrate longer activity hours during the day and week, without typical breaks in weekends, that random, or more normal, profiles show.
\item They are likely (semi)automated accounts, as they rank high in bot scale and regularity of posting.
\squishend
Overall, the profiles we focused-on are small in number compared to the total dataset collected.
This means that consistently toxic misbehavior is still manageable within a popular OSN such as Twitter.
OSN admins can deploy methods like ours to detect and remove such profiles, who are probable participants in influence operations of social discourse. 

\textbf{Future Work:}
We plan to scrutinize further the phenomenon of consistent and highly toxic misbehavior based on features of Twitter profiles such as their self-declared location, and further attempt to automatically detect such profiles using ML classifiers, and across different platforms.

\section*{Acknowledgements}

This work was partially supported by the Macquarie University Cybersecurity Hub (MQCHUB) and the EU H2020 Research and Innovation programme. Hina Qayyum was supported by Macquarie University Domestic High Degree Research Scholarship Program. Nicolas Kourtellis was  partially supported during this project from the EU H2020 Research and Innovation programme under grant agreement No 830927 (Concordia). Any opinions, findings, and conclusions or recommendations expressed in this material are those of the authors or originators and do not necessarily reflect the views of the MQCHUB or of the EU H2020 Research and Innovation program.



\bibliography{reference}

\begin{thebibliography}{32}
\providecommand{\natexlab}[1]{#1}
\providecommand{\url}[1]{\texttt{#1}}
\providecommand{\urlprefix}{URL }
\expandafter\ifx\csname urlstyle\endcsname\relax
  \providecommand{\doi}[1]{doi:\discretionary{}{}{}#1}\else
  \providecommand{\doi}{doi:\discretionary{}{}{}\begingroup
  \urlstyle{rm}\Url}\fi

\bibitem[{Alizadeh et~al.(2020)Alizadeh, Shapiro, Buntain, and
  Tucker}]{doi:10.1126/sciadv.abb5824}
Alizadeh, M.; Shapiro, J.~N.; Buntain, C.; and Tucker, J.~A. 2020.
\newblock Content-based features predict social media influence operations.
\newblock \emph{Science Advances} 6(30).

\bibitem[{Anonymous et~al.(2020)Anonymous, Niaki, Hoang, Gill, and
  Houmansadr}]{Triplet20}
Anonymous; Niaki, A.~A.; Hoang, N.~P.; Gill, P.; and Houmansadr, A. 2020.
\newblock Triplet Censors: Demystifying Great Firewall{\textquoteright}s {DNS}
  Censorship Behavior.
\newblock In \emph{{FOCI}}.

\bibitem[{Blei, Ng, and Jordan(2003)}]{10.5555/944919.944937}
Blei, D.~M.; Ng, A.~Y.; and Jordan, M.~I. 2003.
\newblock Latent Dirichlet Allocation.
\newblock \emph{the Journal of machine Learning research} .

\bibitem[{{Dhungana Sainju} et~al.(2021){Dhungana Sainju}, Mishra, Kuffour, and
  Young}]{DHUNGANASAINJU2021106735}
{Dhungana Sainju}, K.; Mishra, N.; Kuffour, A.; and Young, L. 2021.
\newblock Bullying discourse on Twitter: An examination of bully-related tweets
  using supervised machine learning.
\newblock \emph{CHB} .

\bibitem[{ElSherief et~al.(2018)ElSherief, Nilizadeh, Nguyen, Vigna, and
  Belding}]{elsherief2018peer}
ElSherief, M.; Nilizadeh, S.; Nguyen, D.; Vigna, G.; and Belding, E. 2018.
\newblock Peer to peer hate: Hate speech instigators and their targets.
\newblock In \emph{{ICWSM}}.

\bibitem[{Establishment(2019)}]{FFI-RAPPORT}
Establishment, N. D.~R. 2019.
\newblock Social network centric warfare - understanding influence operations
  in social media.
\newblock Technical report.

\bibitem[{Fernquist et~al.(2019)Fernquist, Svenonius, Kaati, and
  Johansson}]{9108896}
Fernquist, J.; Svenonius, O.; Kaati, L.; and Johansson, F. 2019.
\newblock Extracting Account Attributes for Analyzing Influence on Twitter.
\newblock In \emph{{EISIC}}.

\bibitem[{Flesch(1948)}]{flesch1948new}
Flesch, R. 1948.
\newblock A new readability yardstick.
\newblock \emph{Journal of {AP}} .

\bibitem[{Founta et~al.(2018)Founta, Djouvas, Chatzakou, Leontiadis, Blackburn,
  Stringhini, Vakali, Sirivianos, and Kourtellis}]{founta2018large}
Founta, A.-M.; Djouvas, C.; Chatzakou, D.; Leontiadis, I.; Blackburn, J.;
  Stringhini, G.; Vakali, A.; Sirivianos, M.; and Kourtellis, N. 2018.
\newblock Large Scale Crowdsourcing and Characterization of Twitter Abusive
  Behavior.
\newblock In \emph{AAAI ICWSM}.

\bibitem[{Gini(1912)}]{gini1912variabilita}
Gini, C. 1912.
\newblock Variabilit{\`a} e mutabilit{\`a}.
\newblock \emph{Reprinted in MMS} .

\bibitem[{Gomez et~al.(2019)Gomez, Gibert, Gomez, and
  Karatzas}]{gomez2019exploring}
Gomez, R.; Gibert, J.; Gomez, L.; and Karatzas, D. 2019.
\newblock Exploring Hate Speech Detection in Multimodal Publications.

\bibitem[{Google(2021)}]{perspective}
Google. 2021.
\newblock Perspective API - Using machine learning to reduce toxicity online.
\newblock \url{https://www.perspectiveapi.com/}.

\bibitem[{Hosseini et~al.(2017)Hosseini, Kannan, Zhang, and
  Poovendran}]{hosseini2017deceiving}
Hosseini, H.; Kannan, S.; Zhang, B.; and Poovendran, R. 2017.
\newblock Deceiving Google's Perspective API Built for Detecting Toxic
  Comments.

\bibitem[{Ikeda et~al.(2013)Ikeda, Hattori, Ono, Asoh, and
  Higashino}]{IKEDA201335}
Ikeda, K.; Hattori, G.; Ono, C.; Asoh, H.; and Higashino, T. 2013.
\newblock Twitter user profiling based on text and community mining for market
  analysis.
\newblock \emph{Knowledge-Based Systems} .

\bibitem[{{I}nc.(2021)}]{Fortiguard}
{I}nc., F. 2021.
\newblock {Web Filter Categories}.
\newblock \url{https://fortiguard.com/webfilter/categories}.
\newblock Online; accessed 15-October-2021.

\bibitem[{Jain et~al.(2018)Jain, Brown, Chen, Neaton, Baidas, Dong, Gu, and
  Artan}]{jain2018adversarial}
Jain, E.; Brown, S.; Chen, J.; Neaton, E.; Baidas, M.; Dong, Z.; Gu, H.; and
  Artan, N.~S. 2018.
\newblock Adversarial Text Generation for Google's Perspective API.
\newblock In \emph{CSCI}.

\bibitem[{Jha and Mamidi(2017)}]{jha-mamidi-2017-compliment}
Jha, A.; and Mamidi, R. 2017.
\newblock When does a compliment become sexist? Analysis and classification of
  ambivalent sexism using twitter data.
\newblock In \emph{NLP}.

\bibitem[{Jhaver et~al.(2021)Jhaver, Boylston, Yang, and
  Bruckman}]{10.1145/3479525}
Jhaver, S.; Boylston, C.; Yang, D.; and Bruckman, A. 2021.
\newblock Evaluating the Effectiveness of Deplatforming as a Moderation
  Strategy on Twitter.
\newblock \emph{HCI} .

\bibitem[{Kaggle(2020)}]{kaggle:metoomovement}
Kaggle. 2020.
\newblock Hatred on Twitter During MeToo Movement - Kaggle.
\newblock
  \url{https://www.kaggle.com/rahulgoel1106/hatred-on-twitter-during-metoo-movement}.

\bibitem[{Neethu and Rajasree(2013)}]{6726818}
Neethu, M.~S.; and Rajasree, R. 2013.
\newblock Sentiment analysis in twitter using machine learning techniques.
\newblock In \emph{ICCCNT}.

\bibitem[{Pacheco et~al.(2021)Pacheco, Hui, Torres-Lugo, Truong, Flammini, and
  Menczer}]{pacheco2021uncovering}
Pacheco, D.; Hui, P.-M.; Torres-Lugo, C.; Truong, B.~T.; Flammini, A.; and
  Menczer, F. 2021.
\newblock Uncovering Coordinated Networks on Social Media: Methods and Case
  Studies.

\bibitem[{Pennington, Socher, and Manning(2014)}]{pennington-etal-2014-glove}
Pennington, J.; Socher, R.; and Manning, C. 2014.
\newblock {G}lo{V}e: Global Vectors for Word Representation.
\newblock In \emph{{EMNLP}}.

\bibitem[{{Pew Research Center}(2018)}]{PewResearchCenter2018}
{Pew Research Center}. 2018.
\newblock Bots in the Twittersphere.
\newblock
  \url{https://www.pewresearch.org/internet/2018/04/09/bots-in-the-twittersphere/}.

\bibitem[{Ribeiro et~al.(2018)Ribeiro, Calais, Santos, Almeida, and
  au2}]{ribeiro2018like}
Ribeiro, M.~H.; Calais, P.~H.; Santos, Y.~A.; Almeida, V. A.~F.; and au2, W.
  M.~J. 2018.
\newblock "Like Sheep Among Wolves": Characterizing Hateful Users on Twitter.
\newblock In \emph{{MWSDM}}.

\bibitem[{Rivers and Lewis(2014)}]{rivers2014ethical}
Rivers, C.~M.; and Lewis, B.~L. 2014.
\newblock Ethical research standards in a world of big data.
\newblock \emph{F1000Research} 3.

\bibitem[{Sayyadiharikandeh et~al.(2020)Sayyadiharikandeh, Varol, Yang,
  Flammini, and Menczer}]{botometer}
Sayyadiharikandeh, M.; Varol, O.; Yang, K.-C.; Flammini, A.; and Menczer, F.
  2020.
\newblock Detection of novel social bots by ensembles of specialized
  classifiers.
\newblock In \emph{{ICIKM}}.

\bibitem[{Senter and Smith(1967)}]{senter1967automated}
Senter, R.; and Smith, E.~A. 1967.
\newblock Automated readability index.
\newblock Technical report, {AMRL}.

\bibitem[{Sphinx and Theme(2021)}]{NLTK}
Sphinx; and Theme, N. 2021.
\newblock Natural LAnguage Processing Tool Kit.
\newblock \url{https://www.nltk.org/api/nltk.html}.

\bibitem[{Twitter(2021)}]{twitterAPI}
Twitter. 2021.
\newblock Twitter API Documentation.
\newblock \url{https://developer.twitter.com/en/docs/twitter-api}.

\bibitem[{Waseem(2016)}]{waseem-2016-racist}
Waseem, Z. 2016.
\newblock Are You a Racist or Am {I} Seeing Things? Annotator Influence on Hate
  Speech Detection on {T}witter.
\newblock In \emph{NLP}.

\bibitem[{Waseem and Hovy(2016)}]{waseem-hovy-2016-hateful}
Waseem, Z.; and Hovy, D. 2016.
\newblock Hateful Symbols or Hateful People? Predictive Features for Hate
  Speech Detection on {T}witter.
\newblock In \emph{{NAACL} Student Research Workshop}.

\bibitem[{Yang et~al.(2019)Yang, Varol, Davis, Ferrara, Flammini, and
  Menczer}]{yang2019arming}
Yang, K.-C.; Varol, O.; Davis, C.~A.; Ferrara, E.; Flammini, A.; and Menczer,
  F. 2019.
\newblock Arming the public with artificial intelligence to counter social
  bots.
\newblock \emph{Human Behavior \& Emerging Technologies} .

\end{thebibliography}

\appendix
\clearpage
\onecolumn
\section{Appendix}

\subsection{Perspective API Validation}

As a sanity check of the scores obtained from this API, we focused on the largest seed dataset, i.e., the ICWSM 2018~\cite{founta2018large}, and cross-correlated the scores from the API models with the annotations assigned to the tweet of a user based on the tweet content.
This dataset includes single tweets from 98.3K users, of which 4,940, 13,690, 27,094, and 52,652 were labeled as `hateful', `spam', `abusive', and `normal', respectively. We were able to retrieve tweets from the timelines of a total 39,344 users present in this dataset.
To validate that fact that pre-trained models of Perspective API's are producing stable outputs, we used the API models for ``Toxicity'',``Severe Toxicity'',``Identity Attack'',``Insult'',``Inflammatory'' and ``Threat'', because the definitions of these API scores are closest to the annotation effort from~\cite{founta2018large}.

\begin{figure}[H]
 \centering
 \vspace{-0.5mm}
 \includegraphics[width=0.50\columnwidth]{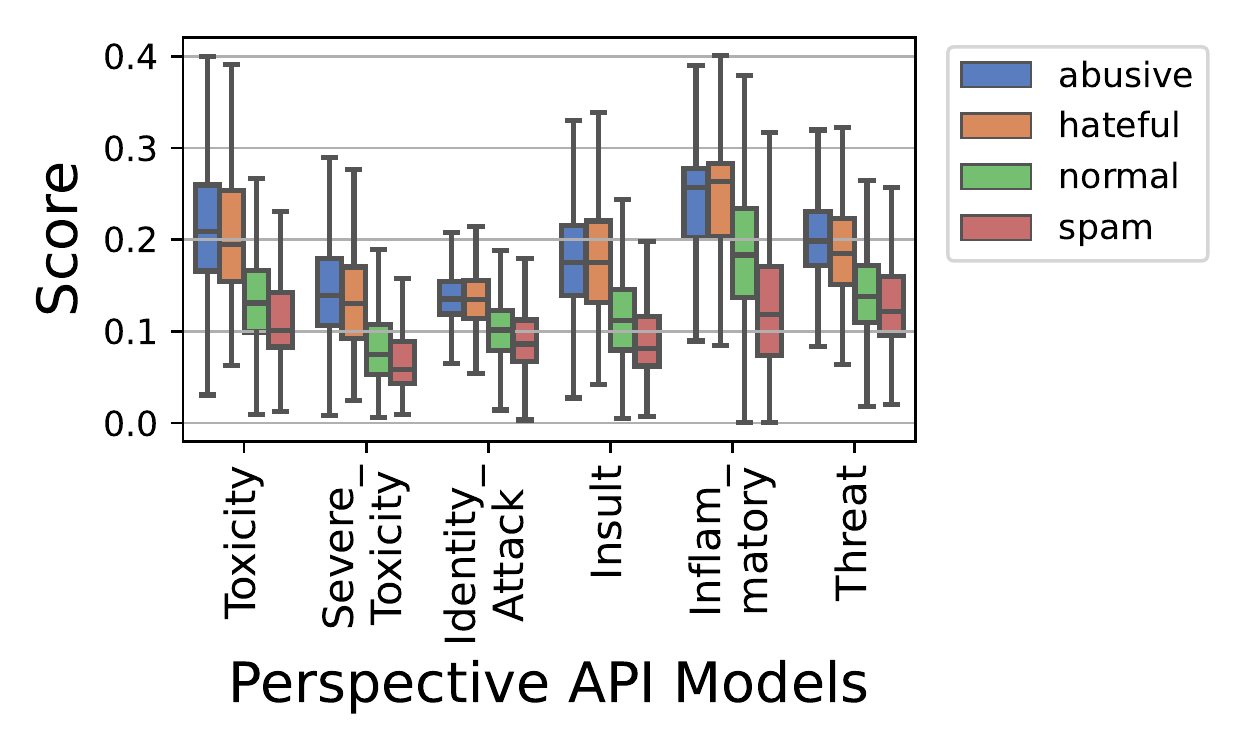}
 \caption{Distribution of scores from six Perspective API models vs. the human annotated labels of tweets.}
 \label{fig:api_validate}
\end{figure}

For all users in the seed dataset, we computed the {\it median} 6 scores mentioned above across all of their tweets. The results of this investigation in Fig.~\ref{fig:api_validate} show the distribution of the API scores, for all four available annotations. Toxicity scores for abusive tweets have a median of 0.2 and highest score are 2.6. Hateful scores have highest value of 2.5. Normal and spam labeled tweets got very low Toxicity scores of 1.8 and 1.5 respectively. In all cases, the distributions of perspective score medians for users labeled ``abusive'' or ``hateful'' are significantly different (p\textless0.01) to those for users labeled ``normal'' or ``spam'', showing consistency between perspective scores and annotated user labels.

\subsection{Focus Profile Selection}

Focus group identification plots of Identity Attack Inflammatory, Insult and Threat scores are presented here. These groups were isolated by the imposing thresholds on respective median scores and Gini index. The details of the process can be found in (\S\ref{sec:focus-users}).
\label{all focus groups}
\begin{figure*}[h!]
\centering
{
\subfloat[Identity Attack]{
\includegraphics[width=0.24\textwidth]{sctr/IDENTITY_ATTACK_gini_median_hina.png}\label{fig}
}
\subfloat[Inflammatory]{
\includegraphics[width=0.24\textwidth]{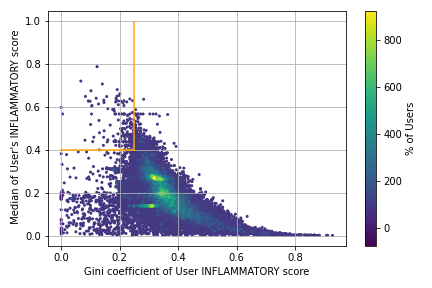}\label{fig}
}
\subfloat[Insult]{
\includegraphics[width=0.24\textwidth]{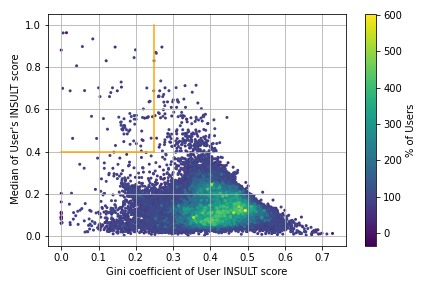}\label{fig:}
}
\subfloat[Threat]{
\includegraphics[width=0.24\textwidth]{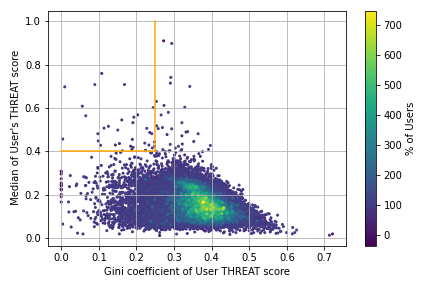}\label{fig:}
}
}
\caption{Focus profiles (bounded by the orange boxes), based on median and Gini coefficient of respective perspective scores of all 3T profiles.}
\label{fig:LABEL}
\end{figure*}
\subsection{URL analysis}

We plotted  the  top  20  categories  of  domains  out  of  all  different  categories  found in focus Identity Attack, focus Inflammatory, focus Insult and focus Threat profiles and their respective random sets of profiles in extension to the analysis performed in  ~\S\ref{sec:url-diversity}.  
\label{url analysis}
\begin{figure*}[t]
\centering
{
\subfloat[Identity Attack]{
\includegraphics[width=0.24\linewidth]{figures/bar_category_IDENTITY_ATTACK_median_xlog-crop.pdf}\label{fig:median:INFLAMMATORY-category}
}
\subfloat[Inflammatory]{
\includegraphics[width=0.24\linewidth]{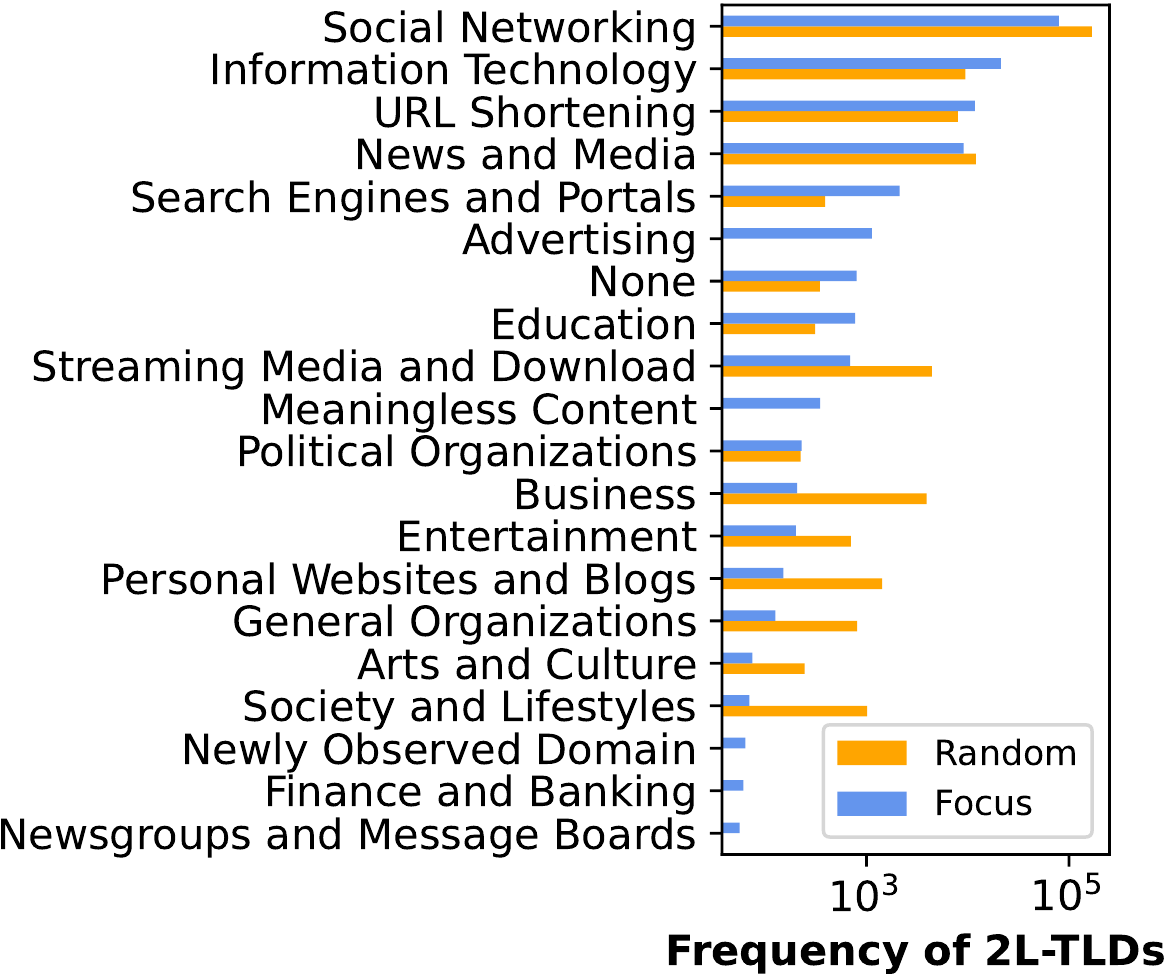}\label{fig:median:INFLAMMATORY-category}
}
\subfloat[Insult]{
\includegraphics[width=0.24\linewidth]{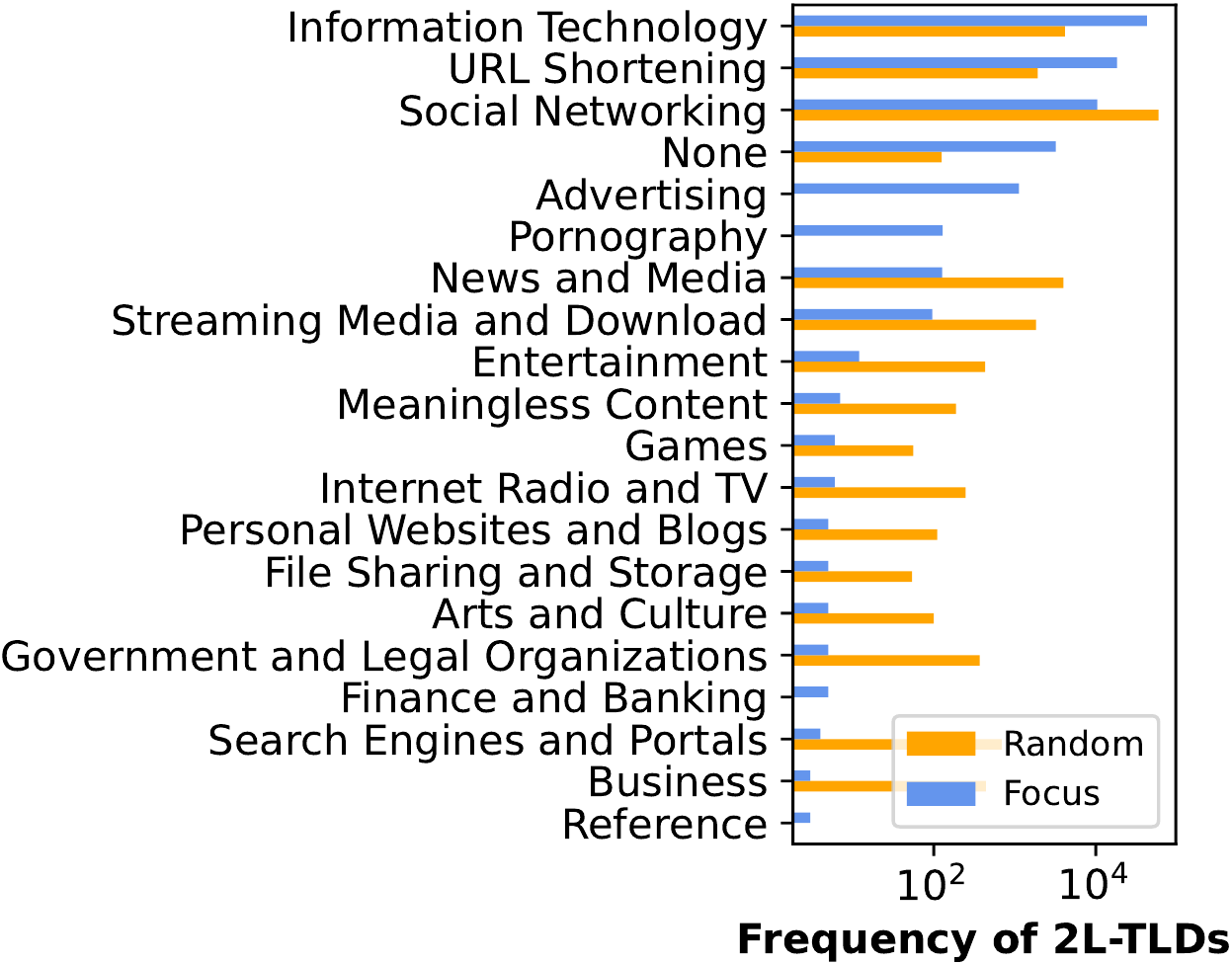}\label{fig:median:INSULT-2-category}
}
\subfloat[Threat]{
\includegraphics[width=0.24\linewidth]{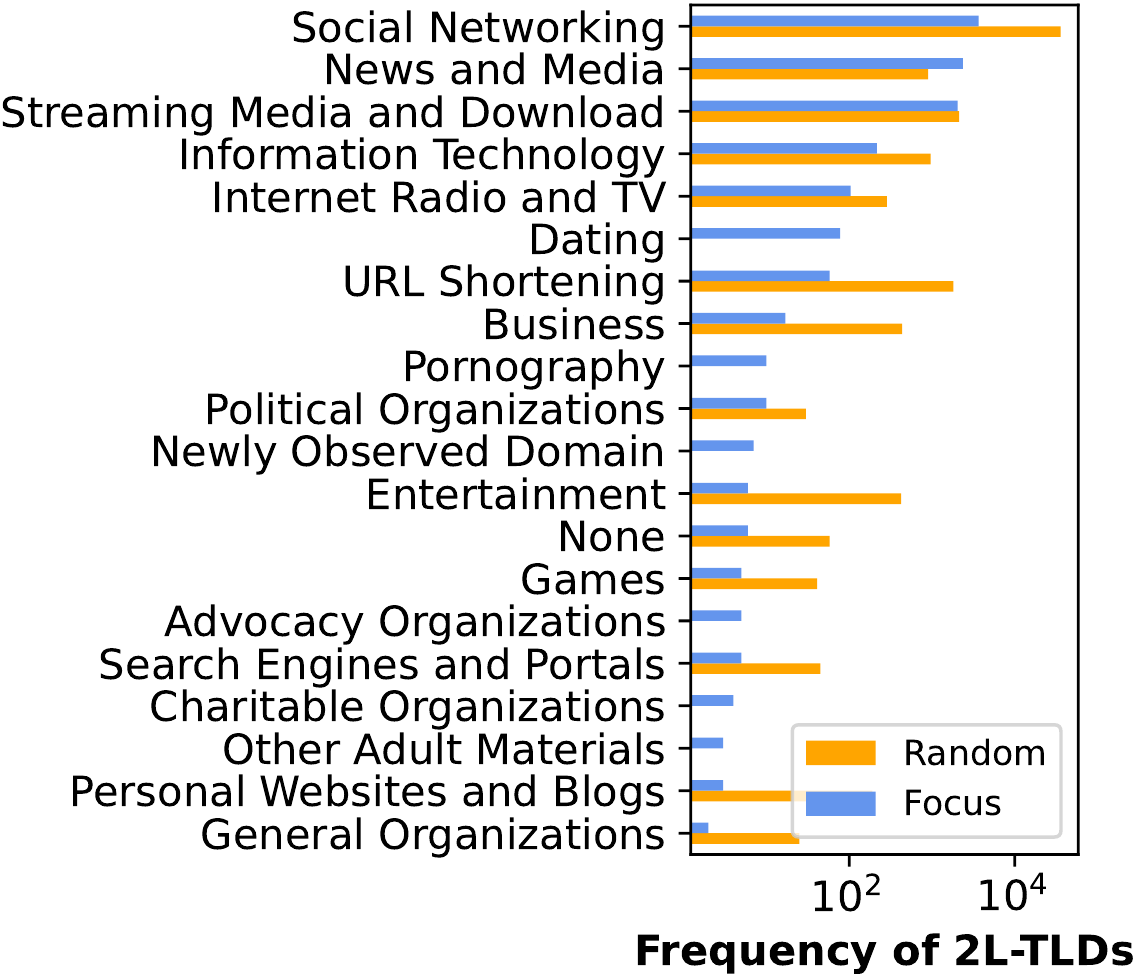}\label{fig:median:THREAT-category}
}
}
\caption{Top 20 domain categories in focus and random Identity Attack, Inflammatory, Insult and Threat sets of profiles. Here, 2L-TLDs refer to second level domains (SLDs). 
``None'' refers to unrated websites whose domain category is unknown to FortiGuard.}
\vspace{-0.9mm}
\label{fig:bar_url_puser:gini_median-category_complete}
\end{figure*}
\newpage
\subsection{Coordination and Time Analysis}
\begin{figure*}[h!]
    \vspace{-0.5mm}
    \begin{subfigure}[t]{0.24\linewidth}
            \includegraphics[width=\textwidth]{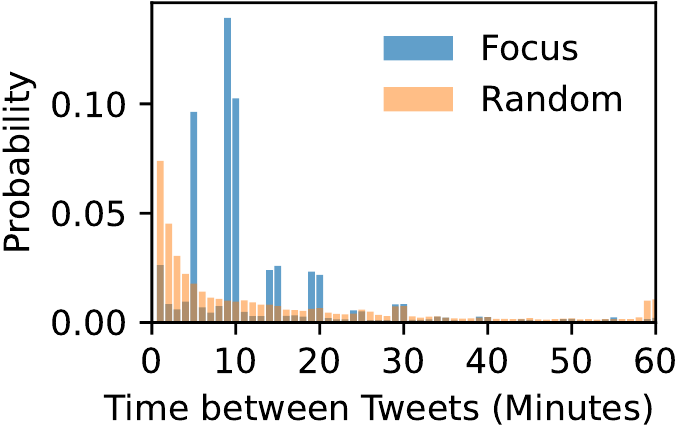}
            \caption{Inter-tweet intervals}  \label{fig:}
    \end{subfigure}\hfill
      \begin{subfigure}[t]{0.24\linewidth}
            \includegraphics[width=\textwidth]{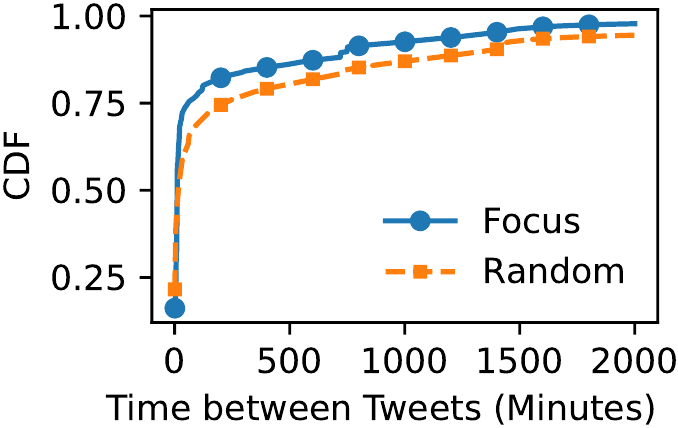}
            \caption{Inter-tweet intervals}
    \label{fig:}
    \end{subfigure}
    \hfill
    \begin{subfigure}[t]{0.24\linewidth}
        \includegraphics[width=\textwidth]{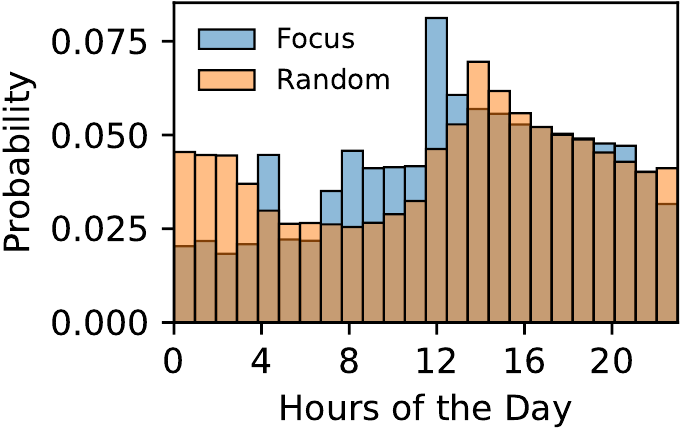}
            \caption{Tweets per hour}
            \label{fig:}
    \end{subfigure}
    \hfill
    \begin{subfigure}[t]{0.24\linewidth}
            \includegraphics[width=\textwidth]{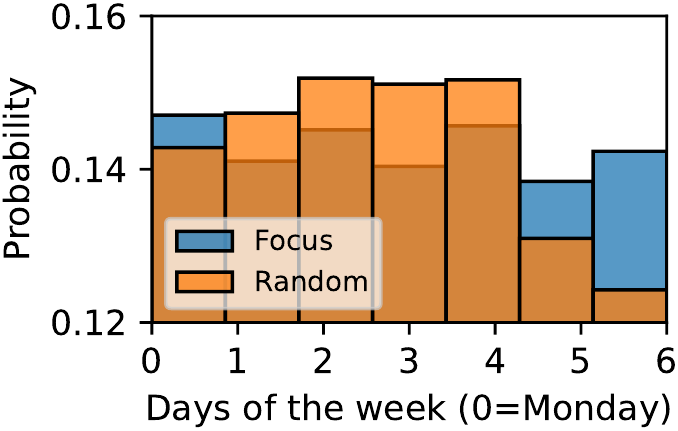}
            \caption{Tweets per day}
            \label{fig:}
    \end{subfigure}
    \caption{Identity Attack}
    \label{fig:day_week_gini_median_IDENTITY_ATTACK}
    \vspace{-2mm}
\end{figure*}

\begin{figure*}[h!]
    \centering
    \begin{subfigure}{0.24\linewidth}
            \includegraphics[width=\textwidth]{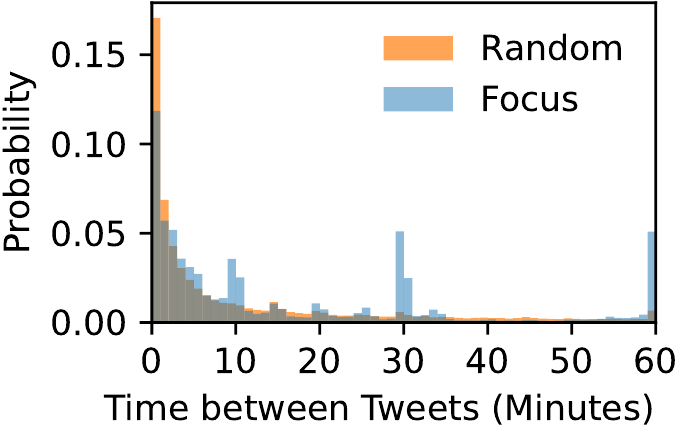}
            \caption{Inter-tweet intervals} 
            \label{fig:}
    \end{subfigure}\hfill
     \begin{subfigure}{0.24\linewidth}
            \includegraphics[width=\textwidth]{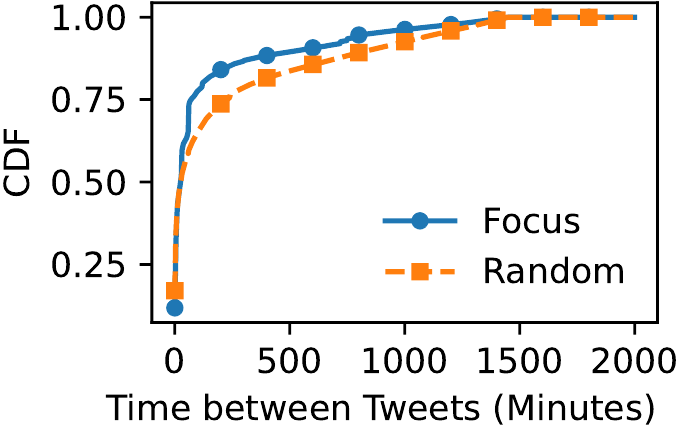}
            \caption{Inter-tweet intervals} 
    \label{fig:}
    \end{subfigure}
    \begin{subfigure}{0.24\linewidth}
        \includegraphics[width=\textwidth]{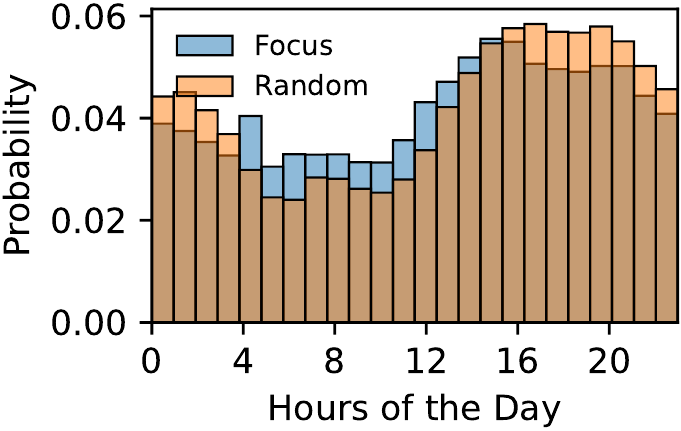}
            \caption{Tweets per hour}
            \label{fig:}
    \end{subfigure}\hfill
    \begin{subfigure}{0.24\linewidth}
            \includegraphics[width=\textwidth]{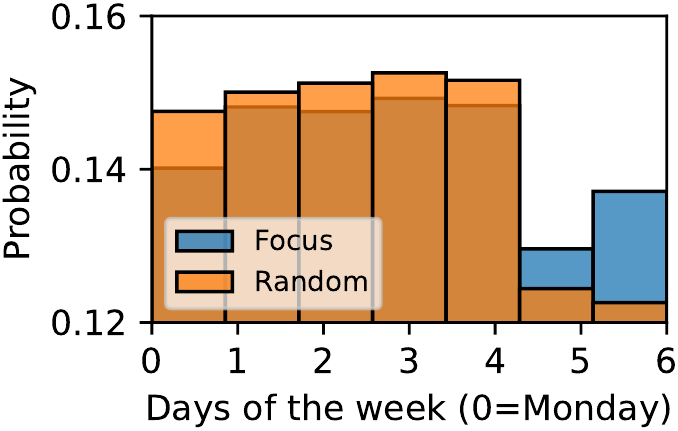}
            \caption{Tweets per day}
            \label{fig:}
    \end{subfigure}
    \caption{Inflammatory}
    \label{fig:day_week_gini_median_INFLAMMATORY}
\end{figure*}

\begin{figure*}[h!]
    \centering
    \begin{subfigure}{0.24\linewidth}
            \includegraphics[width=\textwidth]{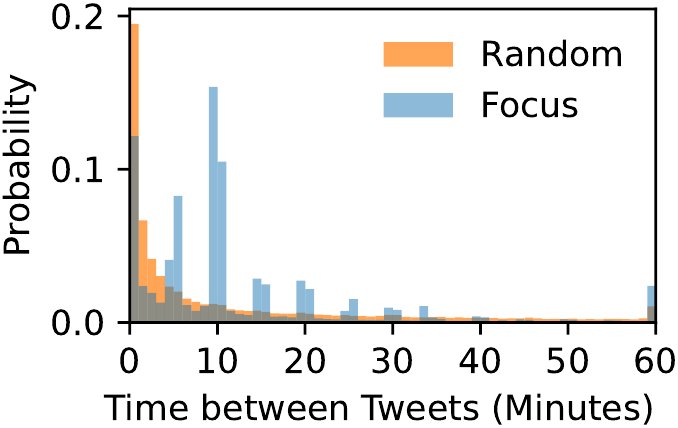}
            \caption{Inter-tweet intervals} 
            \label{fig:}
    \end{subfigure}\hfill
      \begin{subfigure}{0.24\linewidth}
            \includegraphics[width=\textwidth]{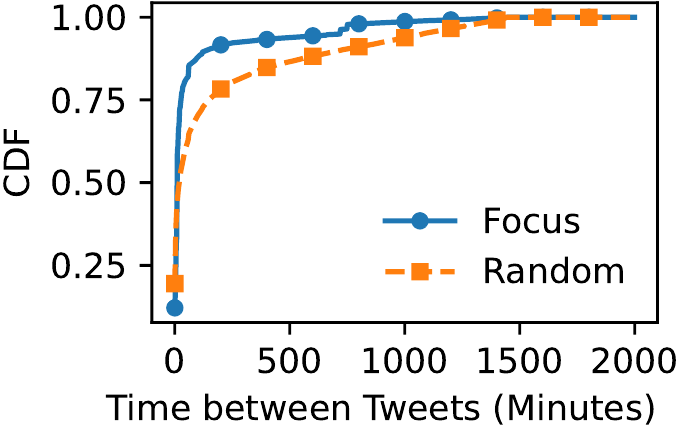}
            \caption{Inter-tweet intervals} 
    \label{fig:}
    \end{subfigure}
    \begin{subfigure}{0.24\linewidth}
        \includegraphics[width=\textwidth]{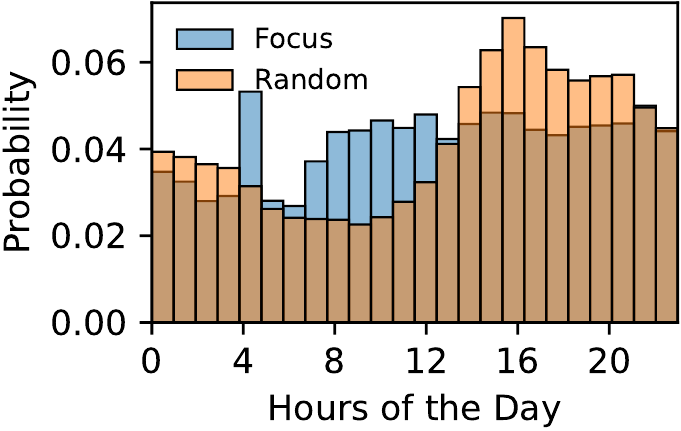}
            \caption{Tweets per hour}
            \label{fig:}
    \end{subfigure}\hfill
    \begin{subfigure}{0.24\linewidth}
            \includegraphics[width=\textwidth]{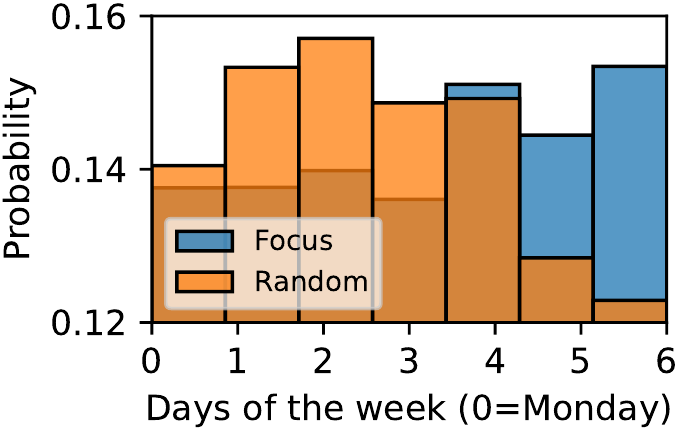}
            \caption{Tweets per day}
            \label{fig:}
    \end{subfigure}
    \caption{Insult}
    \label{fig:day_week_gini_median_INSULT}
\end{figure*}

\begin{figure*}[h!]
    \centering
    \begin{subfigure}{0.24\linewidth}
            \includegraphics[width=\textwidth]{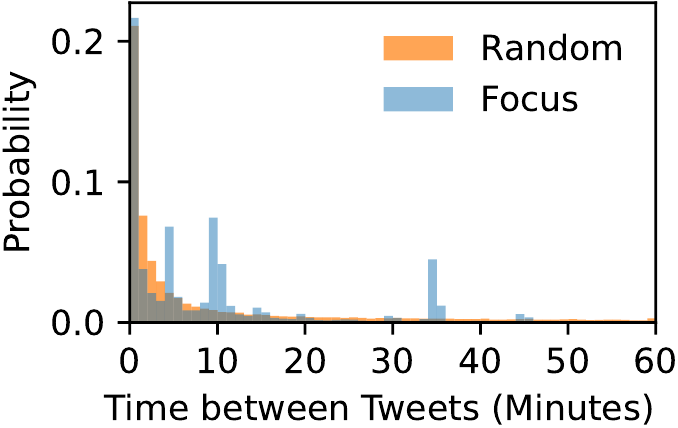}
            \caption{Inter-tweet intervals} 
            \label{fig:}
    \end{subfigure}\hfill
     \begin{subfigure}{0.24\linewidth}
            \includegraphics[width=\textwidth]{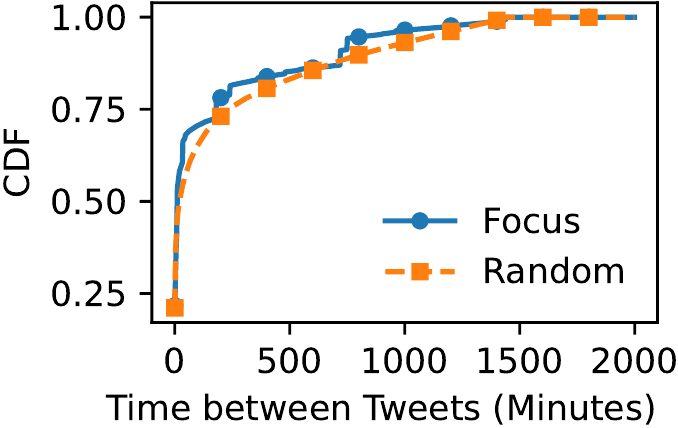}
            \caption{Inter-tweet intervals} 
    \label{fig:}
    \end{subfigure}
    \begin{subfigure}{0.24\linewidth}
        \includegraphics[width=\textwidth]{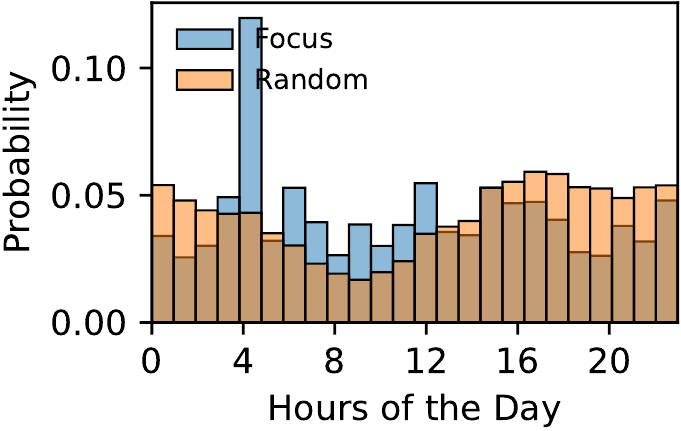}
            \caption{Tweets per hour}
            \label{fig:}
    \end{subfigure}\hfill
    \begin{subfigure}{0.24\linewidth}
            \includegraphics[width=\textwidth]{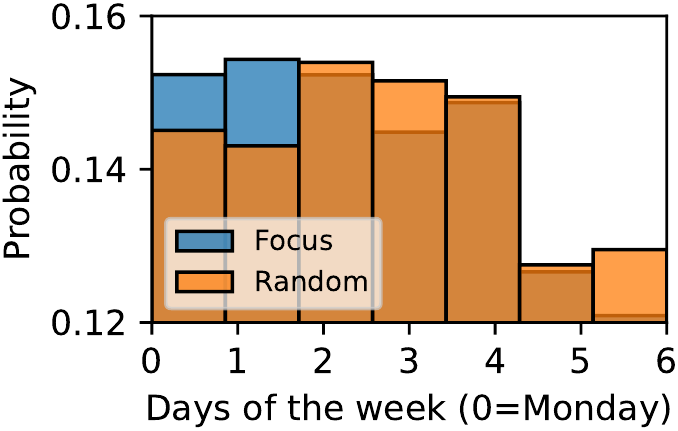}
            \caption{Tweets per day}
            \label{fig:}
    \end{subfigure}
    \caption{Threat}
    \textbf{
    Overview of temporal analysis and coordination analysis: (a) PDF of tweet inter-arrival time;
    (b) CDF of tweet inter-arrival time;
    (c) PDF of tweet posting time during the day;
    (d) PDF of tweet posting time during the week.
}
    \label{fig:day_week_gini_median_THREAT}
    \end{figure*}

Subplots in Figures~\ref{fig:day_week_gini_median_IDENTITY_ATTACK}~\ref{fig:day_week_gini_median_INFLAMMATORY},~\ref{fig:day_week_gini_median_INSULT} and~\ref{fig:day_week_gini_median_THREAT} are shared for focus Identity Attack, Inflammatory, Insult and Threat profiles and the respective random groups in extension to analysis performed in ~\ref{sec:tweeting-patterns}. Sub-plots (a) in these figures show the Probability Distribution Function (PDF) of time between sequential tweets by profiles of focus and random profiles, up to 60 minutes. Sub-plots(b) show the CDF of inter-tweet intervals as an extension to aforementioned PDF.
We also looked into the time of day and day of week at which tweets are posted by focus or random profiles across three types of misbehavior in sub-plots (c) and (d) of all four sets of figures.
\label{A:coordination analysis}

\subsection{Number of URLs and domains per focus and random profile}
\label{sec:no.url/domains}

\begin{figure*}[h!]
\centering
{
\subfloat[Identity Attack]{
\includegraphics[width=0.24\textwidth]{figures/domains_url_CDF_IDENTITY_ATTACK_median.pdf}\label{fig:}
}
\subfloat[Inflammatory]{
\includegraphics[width=0.24\textwidth]{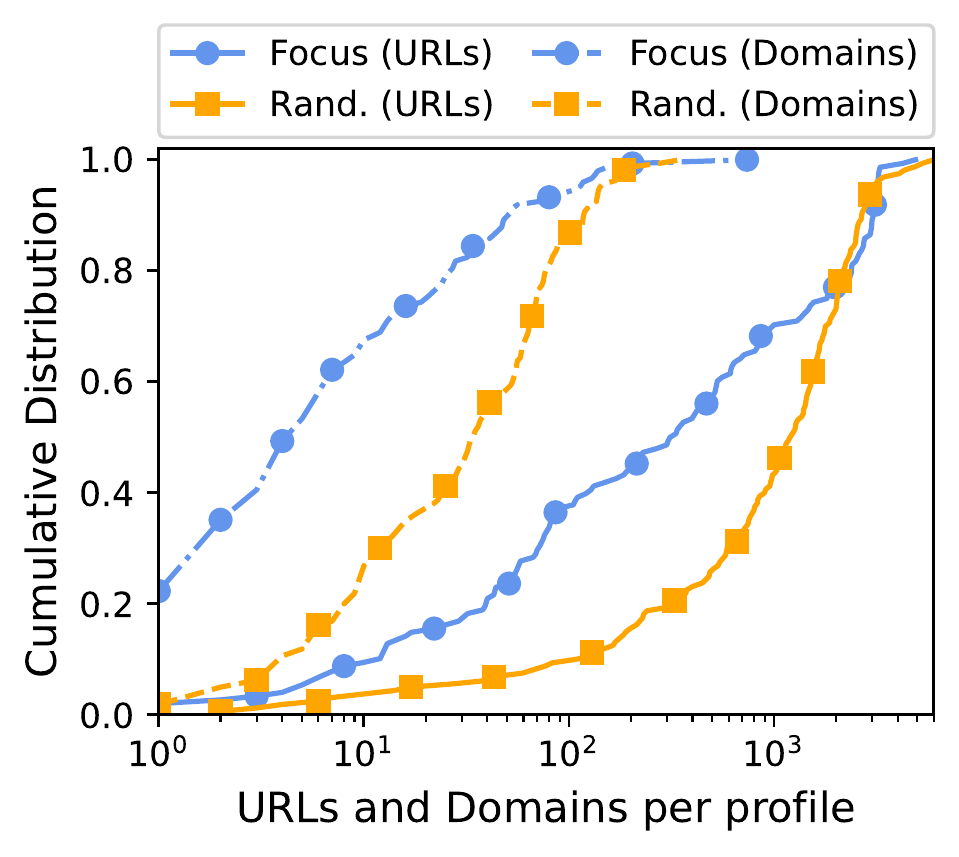}\label{fig:}
}
\subfloat[Insult]{
\includegraphics[width=0.24\textwidth]{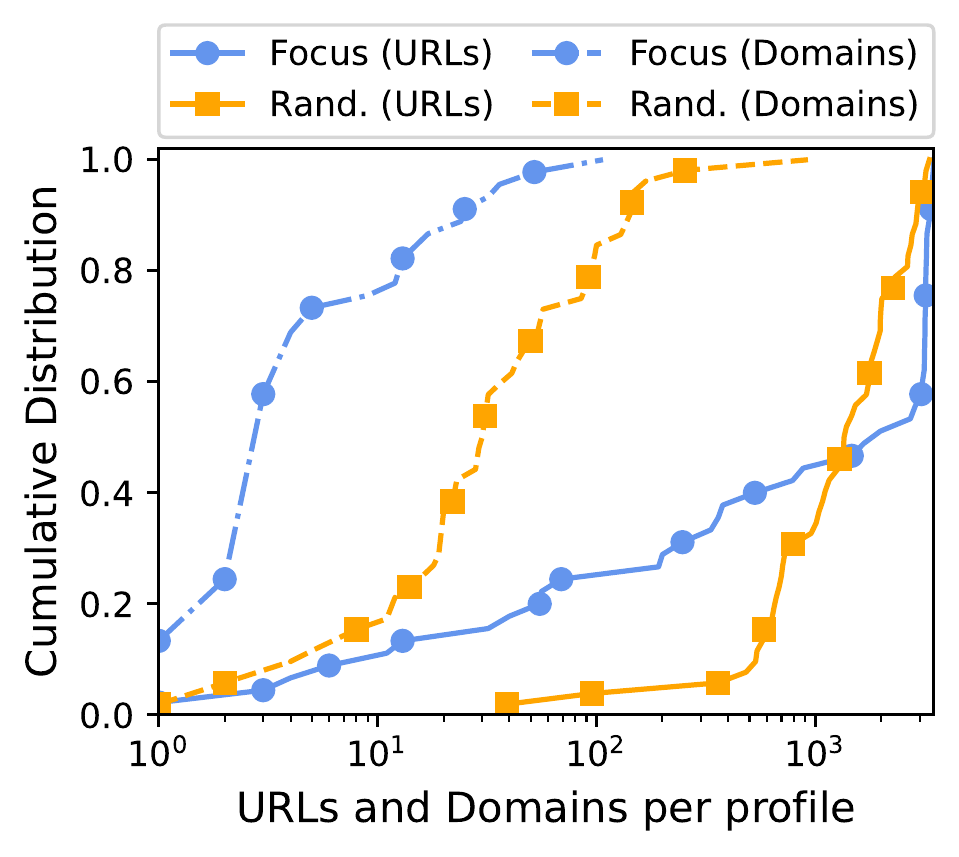}\label{fig:}
}
\subfloat[Threat]{
\includegraphics[width=0.24\textwidth]{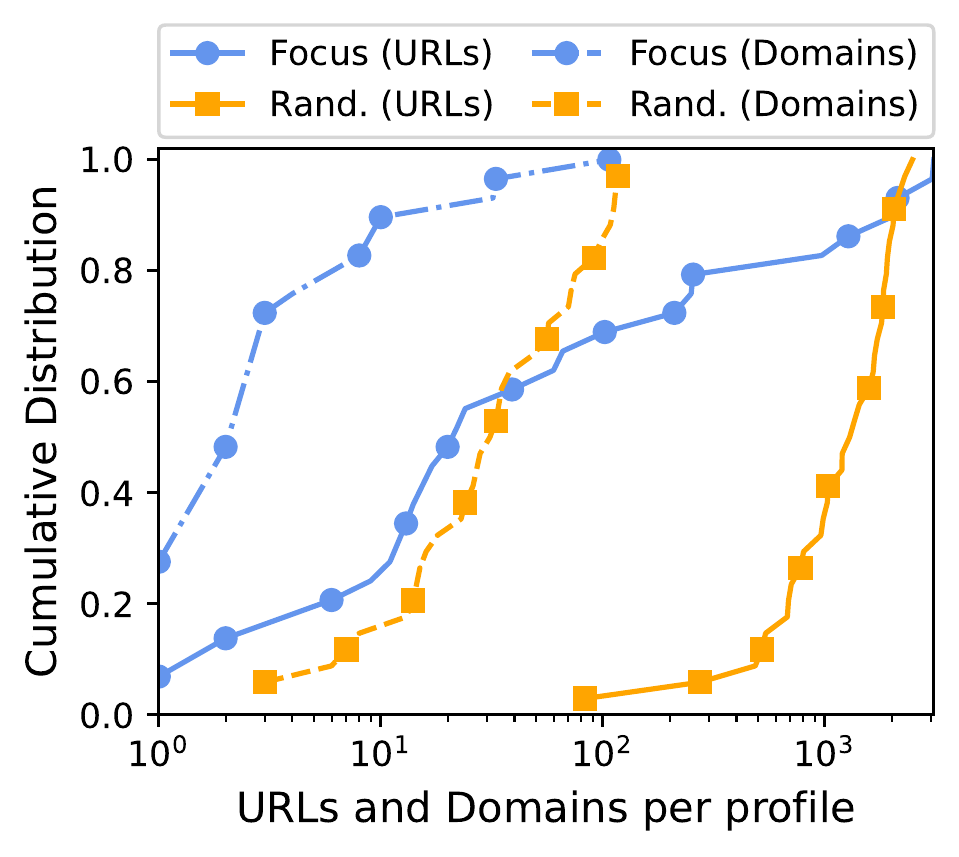}\label{fig:}
}
}
\caption{CDFs of the Number of Unique URLs and Domains in Tweets of Focus and Random Group Profiles}
\label{fig:urls and domains in tweets}
\end{figure*}

In Figure~\ref{fig:urls and domains in tweets} we share the CDFs of the number of unique URLs and referenced domains in tweets of focus and random sets of profiles of all four types of investigated misbehavior such as Identity Attack, inflammatory and such (Section:~\ref{sec:url-diversity}).

\subsection{Number of hashtags per focus and random profile}
\label{sec:no.of hashtags}

\begin{figure*}[h!]
\centering
{
\subfloat[Identity Attack]{
\includegraphics[width=0.24\textwidth]{figures/hashtags_CDF_IDENTITY_ATTACK_median.pdf}\label{fig:}
}
\subfloat[Inflammatory]{
\includegraphics[width=0.24\textwidth]{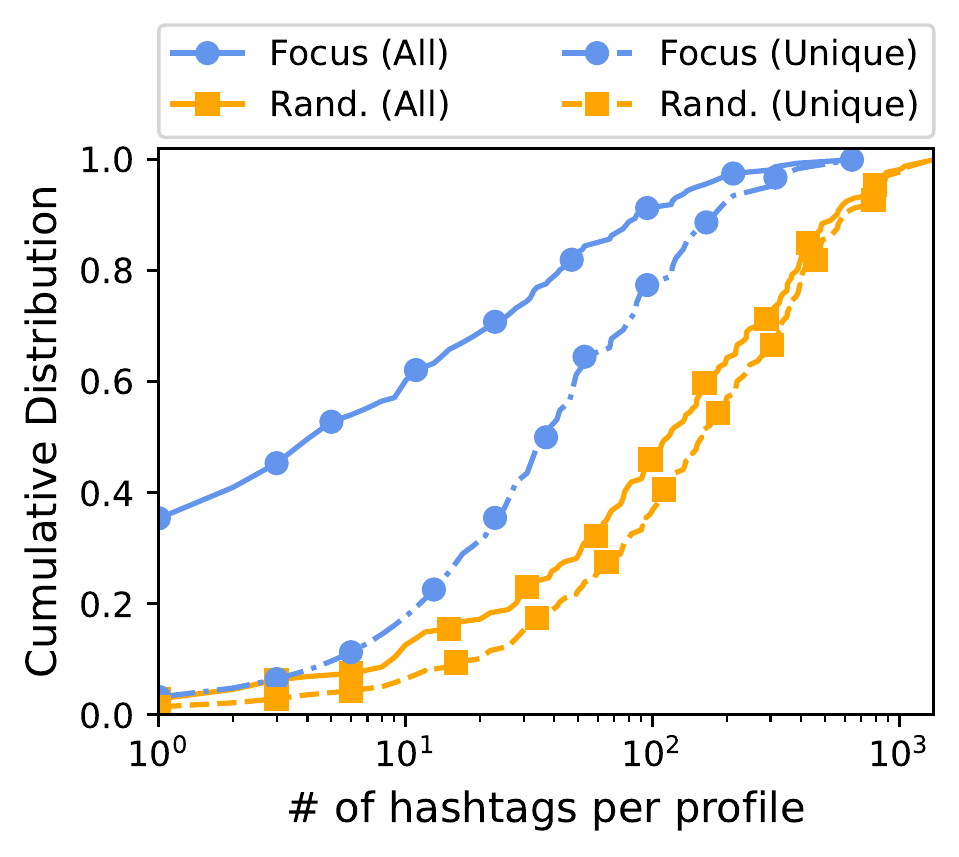}\label{fig:}
}
\subfloat[Insult]{
\includegraphics[width=0.24\textwidth]{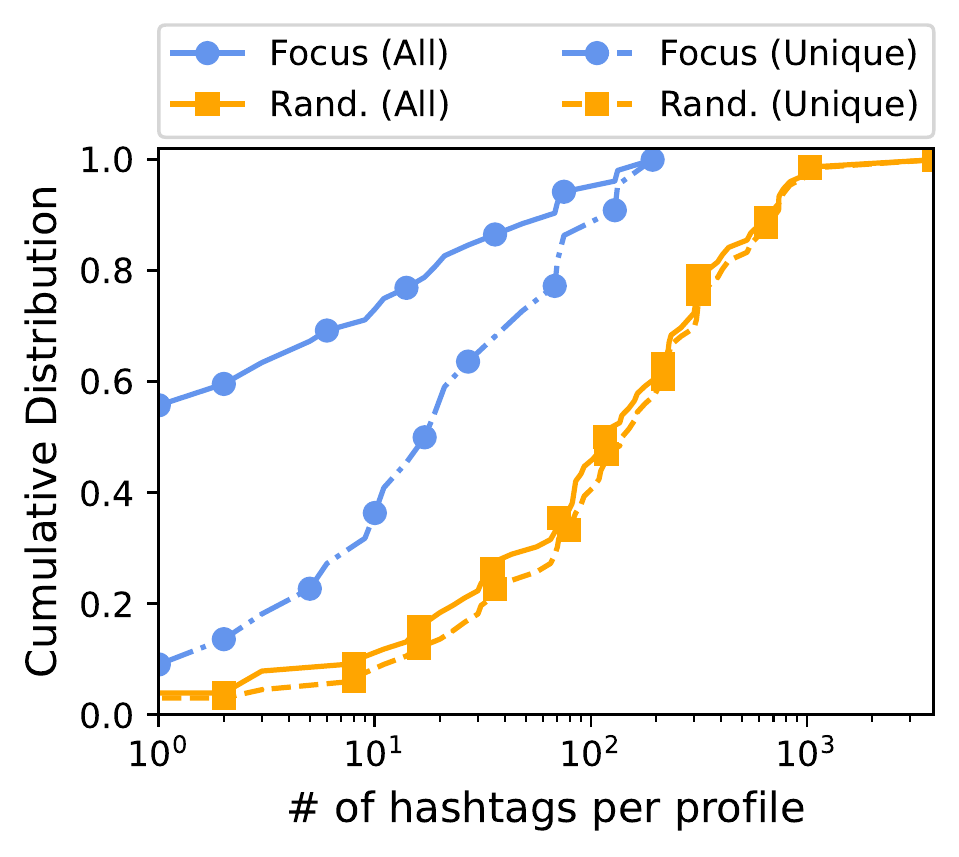}\label{fig:}
}
\subfloat[Threat]{
\includegraphics[width=0.24\textwidth]{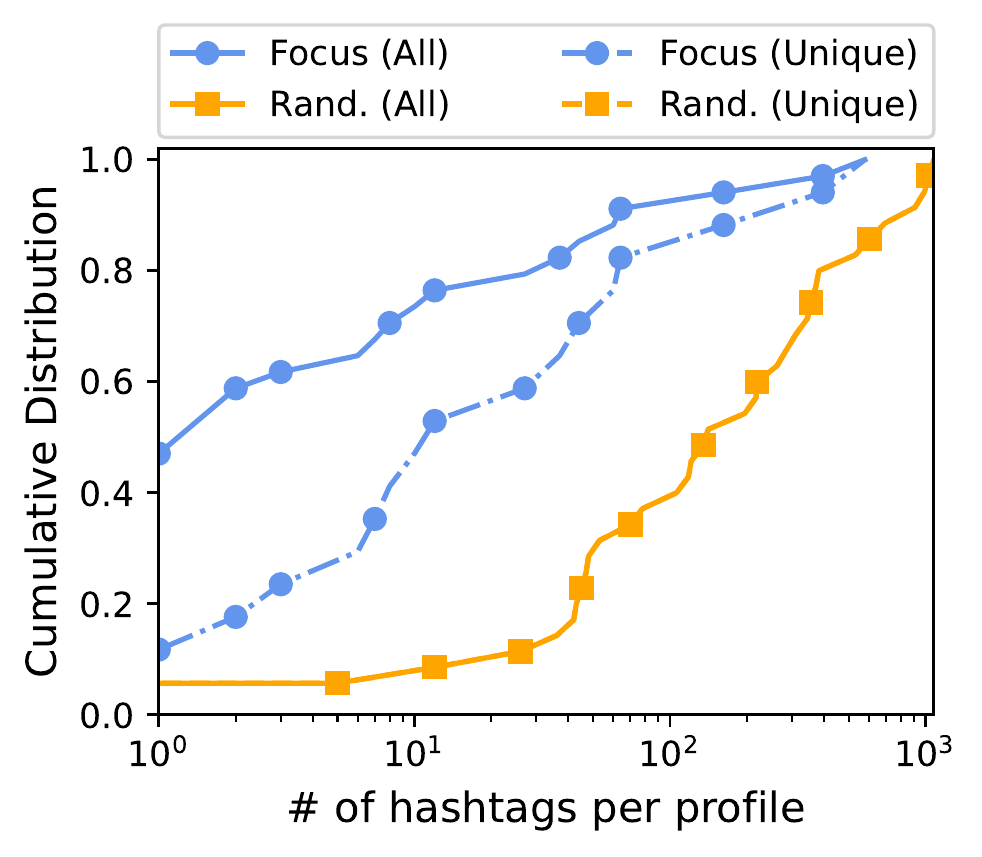}\label{fig:}
}
}
\caption{CDFs of the Number of Unique Hashtags in Tweets of Focus and Random Group Profiles}
\label{fig:number of hashtags in tweets}
\end{figure*}

Figure~\ref{fig:number of hashtags in tweets} represents the CDFs of the number of hashtags shared in tweets of focus and random profiles across all categories of misbehavior including Identity Attack, Inflammatory, Insult and Threat (Section:~\ref{sec:hashtag-diversity}).

\subsection{URL Jaccard Similarity}

\begin{figure*}[!]
\centering
{
\subfloat[Identity Attack]{
\includegraphics[width=0.24\textwidth]{figures/domains_similarity_IDENTITY_ATTACK_median_xlog.pdf}\label{fig:}
}
\subfloat[Inflammatory]{
\includegraphics[width=0.24\textwidth]{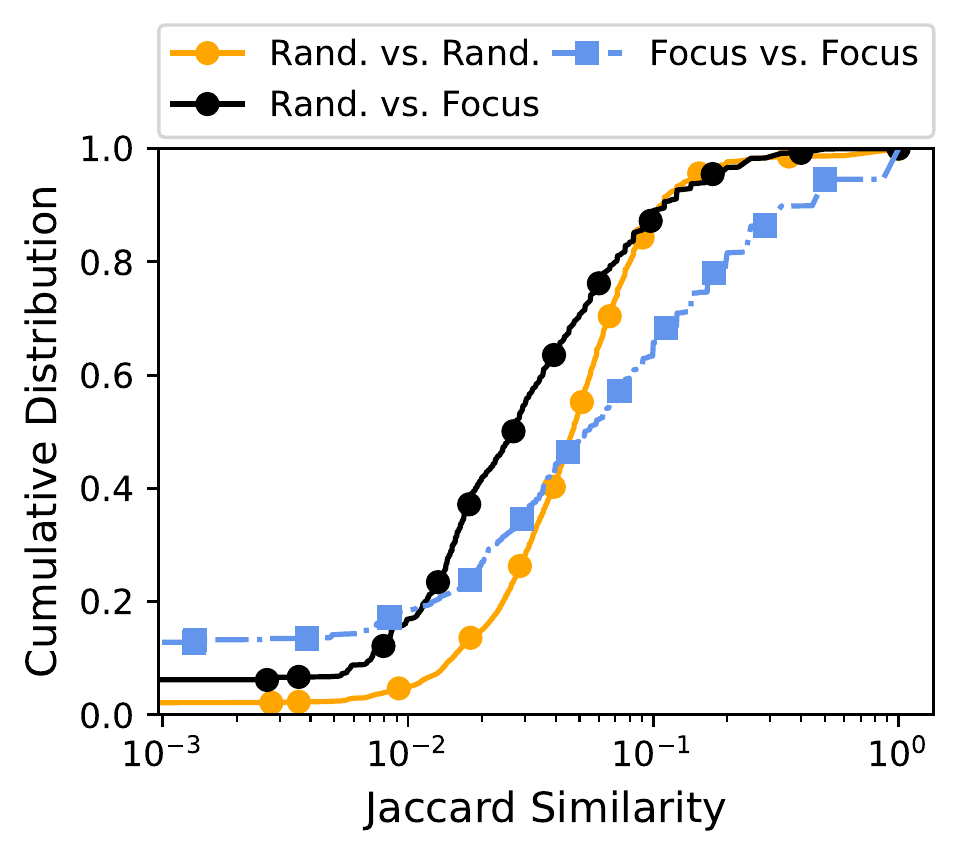}\label{fig:}
}
\subfloat[Insult]{
\includegraphics[width=0.24\textwidth]{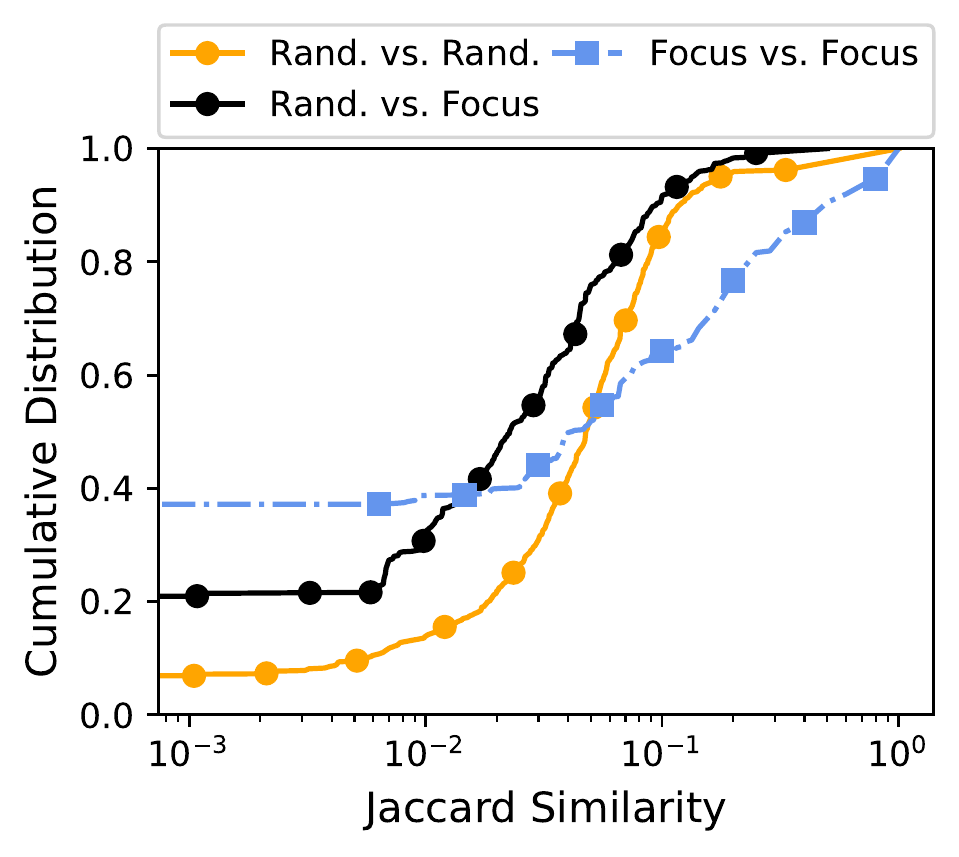}\label{fig:}
}
\subfloat[Threat]{
\includegraphics[width=0.24\textwidth]{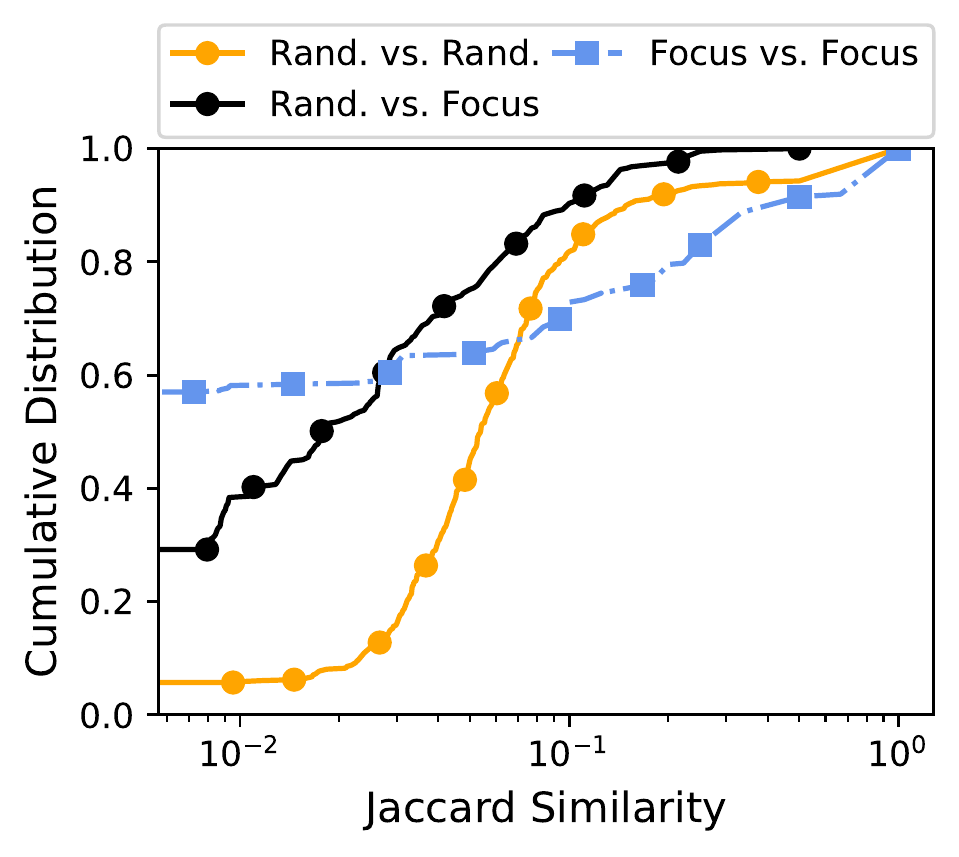}\label{fig:}
}
}
\caption{Domain Similarity among profiles of Focus and Random Profiles}
\label{fig:domain similarity}
\end{figure*}

Figure~\ref{fig:domain similarity} furthers the results of Section ~\ref{sec:similarity-domains}  and Figure ~\S\ref{fig:url_domain_puser:gini_median-dom-simm} for the Perspective scores of Inflammatory, Insult, Threat and their respective random groups. The plots highlight the similarity of domains in focus and random groups and inter-similarity of domains found in tweets of focus and random groups across all four types of misbehavior.

\newpage
\subsection{Hashtag Jaccard Similarity}
\begin{figure}[H]
\centering
{
\subfloat[Identity Attack]{
\includegraphics[width=0.24\textwidth]{figures/hashtags_jaccard_similarity_IDENTITY_ATTACK_median.pdf}\label{fig:}
}
\subfloat[Inflammatory]{
\includegraphics[width=0.24\textwidth]{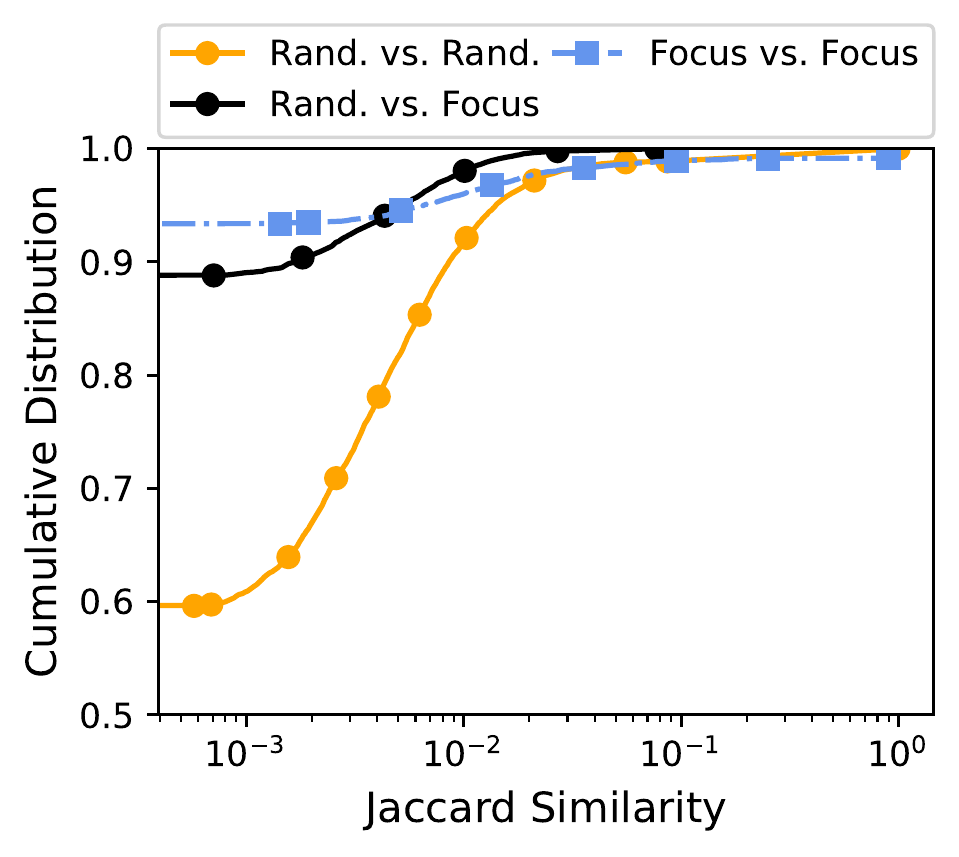}\label{fig:}
}
\subfloat[Insult]{
\includegraphics[width=0.24\textwidth]{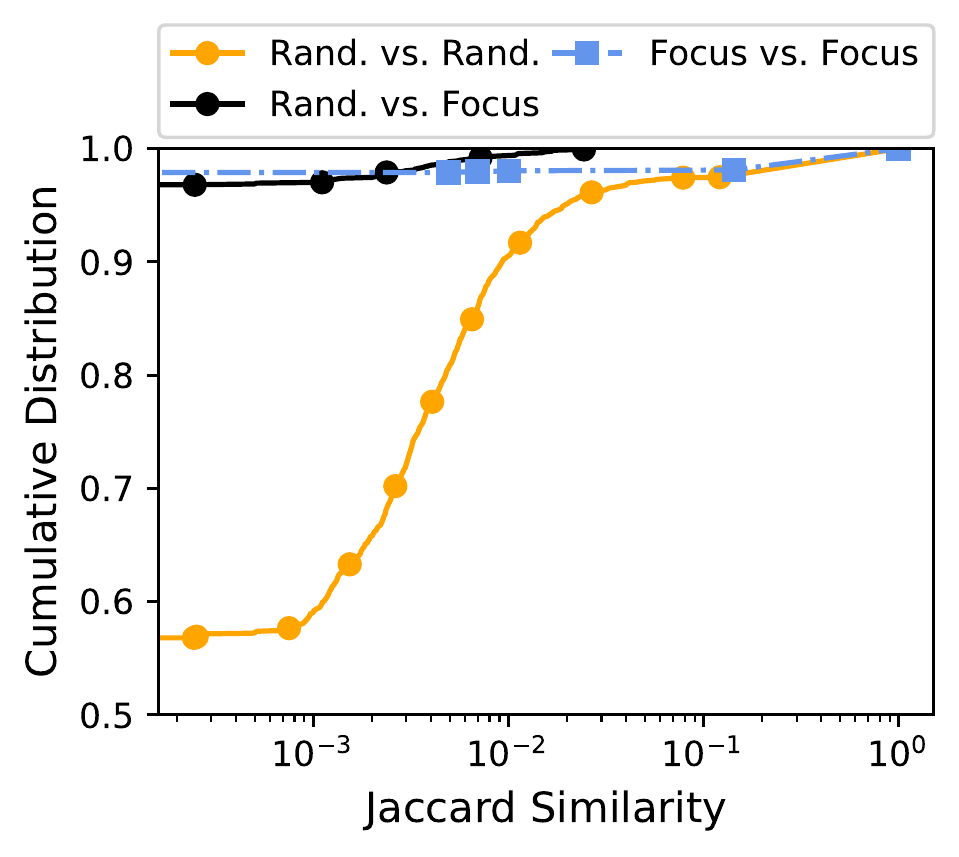}\label{fig:}
}
\subfloat[Threat]{
\includegraphics[width=0.24\textwidth]{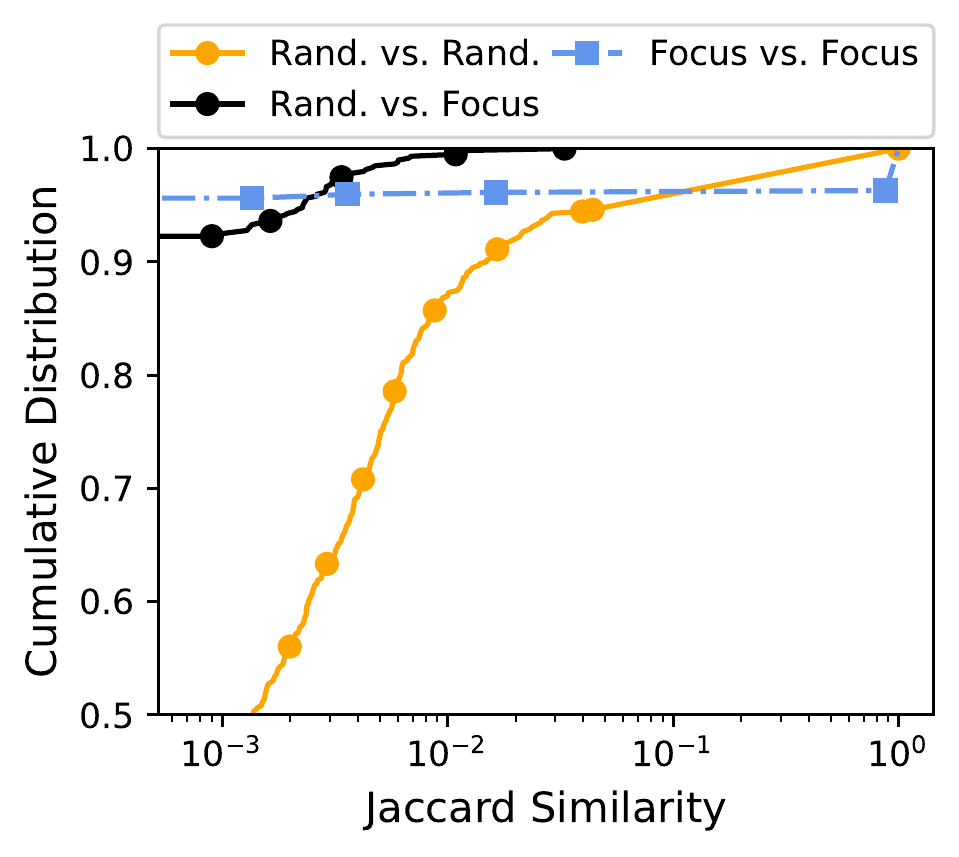}\label{fig:}
}
}
\caption{Jaccard
Similarity among Hashtags of Focus and Random Profiles}
\label{fig:hashtags jaccard similarity}
\end{figure}

These plots in Figure~\ref{fig:hashtags jaccard similarity} are shared to extend the results of Section ~\ref{sec:similarity-domains} for focus and random profiles of Inflammatory, Insult and Threat categories of misbehavior.

\subsection{Hashtag Explanation}

\begin{table*}[hb]
\resizebox{\textwidth}{!}{
\begin{tabular}{|l|l|}
\hline
Hashtag & Explanation \\
\hline
\#TreCru   & One Piece Treasure Cruise, a F2P RPG game   based on the popular manga and anime One Piece                 \\ \hline
\#BDS      & The Palestinian-led BDS movement promotes boycotts, divestments, and economic   sanctions against Israel \\ \hline
\#BREAKING & A hashtag used to represent breaking news.                                                                 \\ \hline
\#BlackLivesMatter & A political and social movement that seeks to highlight racism, discrimination, and inequality experienced by black people. \\ \hline
\#MeToo    & Me Too is a movement against sexual abuse and harassment through public disclosure of   allegations.     \\ \hline
\#MAGA     & Make America Great Again was a campaign slogan leading up to and during   the Trump presidency             \\ \hline
\#trap     & A subgenre of hip-hop music                                                                                \\ \hline
pg3d     & Pixel Gun 3D is a online multiplayer FPS heavily influenced by the pixel art style of Minecraft         \\ \hline
\end{tabular}
}
\caption{Explanations for lesser known hashtags that occur in Table~\ref{tab:top5_htags_gini_median}.}
\label{tab:explain_hashtag}
\end{table*}

\end{document}